\documentclass[journal=jpclcd]{achemso}

\usepackage{amsmath,bm}
\usepackage{amssymb}
\usepackage{mathrsfs}
\usepackage{amsfonts}
\usepackage{graphicx}
\usepackage{siunitx}
\usepackage[version=3]{mhchem}
\usepackage[acronym]{glossaries}

\usepackage{xfrac}

\DeclareSIUnit\angstrom{\text {Å}}
\DeclareSIUnit\debye{\text {D}}

\newcommand{\ke}[1]{\vert #1 \rangle}

\newacronym{cboa}{CBOA}{cavity Born-Oppenheimer approximation}
\newacronym{bo}{BOA}{Born-Oppenheimer approximation}
\newacronym{cbohf}{CBO-HF}{cavity Born-Oppenheimer Hartree-Fock}
\newacronym{scf}{SCF}{self-consistent field}
\newacronym{dse}{DSE}{dipole self-energy}
\newacronym{pes}{PES}{potential energy surface}
\newacronym{cpes}{cPES}{cavity potential energy surface}
\newacronym{vsc}{VSC}{vibrational-strong coupling}
\newacronym{esc}{ESC}{electronic-strong coupling}

\title{Cavity-Born-Oppenheimer Hartree-Fock Ansatz: Light-matter Properties of Strongly Coupled
Molecular Ensembles}

\author{Thomas Schnappinger}
\affiliation{Department of Physics, Stockholm University, AlbaNova University Center, SE-106 91 Stockholm, Sweden}
  
\author{Dominik Sidler}
\affiliation{Max Planck Institute for the Structure and Dynamics of Matter and Center for Free-Electron Laser Science, Luruper Chaussee 149, 22761 Hamburg, Germany}
\alsoaffiliation{The Hamburg Center for Ultrafast Imaging, Luruper Chaussee 149, 22761 Hamburg, Germany}

\author{Michael Ruggenthaler}
\affiliation{Max Planck Institute for the Structure and Dynamics of Matter and Center for Free-Electron Laser Science, Luruper Chaussee 149, 22761 Hamburg, Germany}
\alsoaffiliation{The Hamburg Center for Ultrafast Imaging, Luruper Chaussee 149, 22761 Hamburg, Germany}

\author{Angel Rubio}
  \affiliation{Max Planck Institute for the Structure and Dynamics of Matter and Center for Free-Electron Laser Science, Luruper Chaussee 149, 22761 Hamburg, Germany}
    \alsoaffiliation{The Hamburg Center for Ultrafast Imaging, Luruper Chaussee 149, 22761 Hamburg, Germany}
  \alsoaffiliation{Center for Computational Quantum Physics, Flatiron Institute, 162 5th Avenue, New York, NY 10010, USA}
  \alsoaffiliation{Nano-Bio Spectroscopy Group, University of the Basque Country (UPV/EHU), 20018 San Sebasti\'an, Spain}
  
\author{Markus Kowalewski}
\email{markus.kowalewski@fysik.su.se}
\affiliation{Department of Physics, Stockholm University, AlbaNova University Center, SE-106 91 Stockholm, Sweden}
\begin{document}

\setlength{\fboxrule}{0 pt}
\begin{tocentry}
\includegraphics{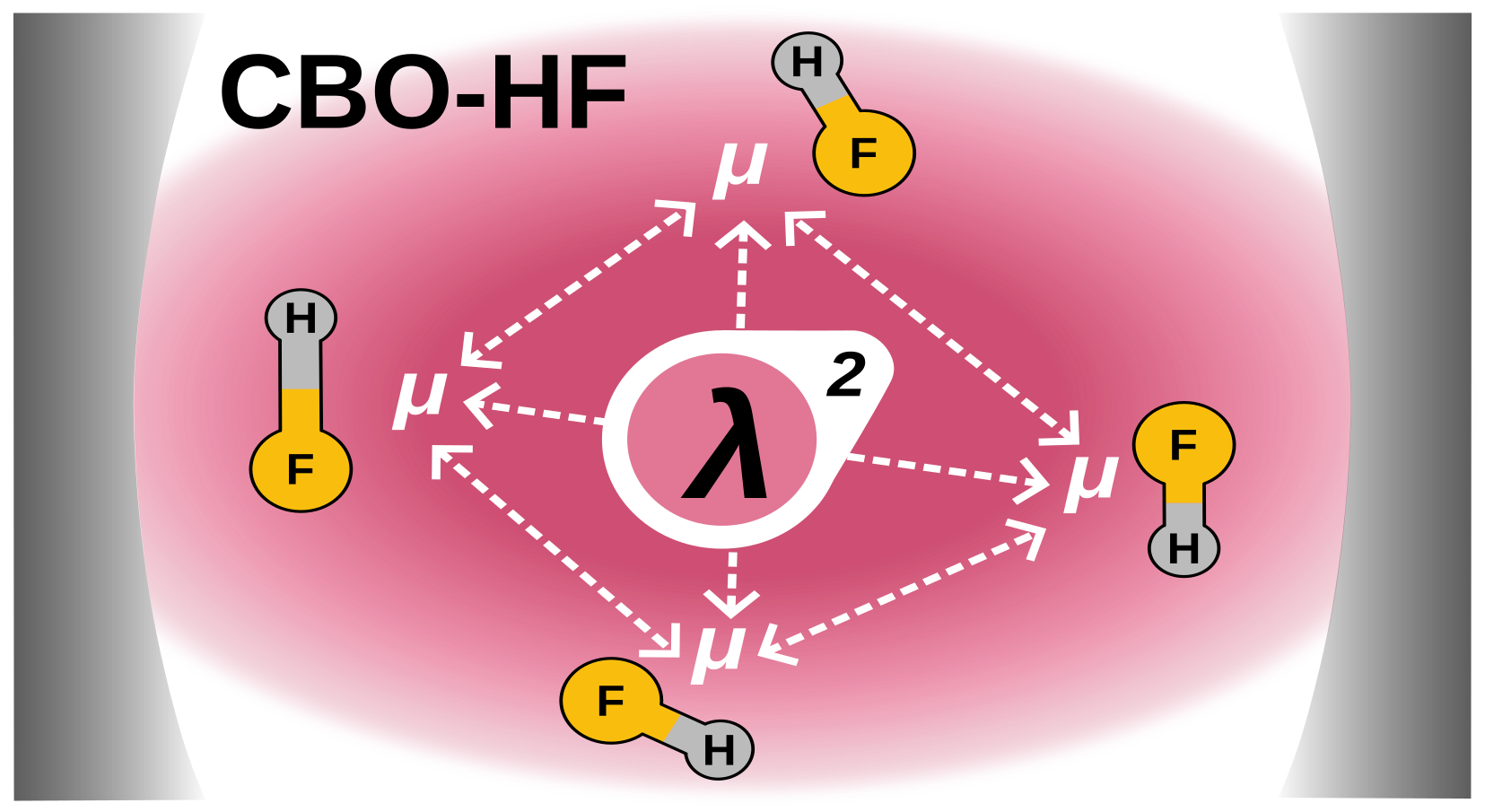}
\end{tocentry}

\begin{abstract}
Experimental studies indicate that optical cavities can affect chemical reactions, through either vibrational or electronic strong coupling and the quantized cavity modes.
However, the current understanding of the interplay between molecules and confined light modes is incomplete. 
Accurate theoretical models, that take into account inter-molecular interactions to describe ensembles, are therefore essential to understand the mechanisms governing polaritonic chemistry.
We present an \textit{ab-initio} Hartree-Fock ansatz in the framework of the cavity Born-Oppenheimer approximation and study molecules strongly interacting with an optical cavity.
This ansatz provides a non-perturbative, self-consistent description of strongly coupled molecular ensembles taking into account the cavity-mediated dipole self-energy contributions. 
To demonstrate the capability of the cavity Born-Oppenheimer Hartree-Fock ansatz, we study the collective effects in ensembles of strongly coupled diatomic hydrogen fluoride molecules.
Our results highlight the importance of the cavity-mediated inter-molecular dipole-dipole interactions, which lead to energetic changes of individual molecules in the coupled ensemble.
\end{abstract}

\maketitle

The strong coupling of light and matter within optical cavities provides an innovative way to alter and design matter properties, making it a rapidly evolving field~\cite{Ebbesen2016-jx,Ruggenthaler2018-ew,Gargiulo2019-cy,Herrera2020-bg,Nagarajan2021-sl,Sidler2022-cg,Li2022-gi,Fregoni2022-op,Dunkelberger2022-oh,Tibben2023-if} at the intersection of chemistry, quantum optics, and materials science. In polaritonic chemistry, depending on whether the quantized cavity modes are coupled via their characteristic frequencies to electronic or vibrational degrees of freedom of molecules, the situation is described as \gls{esc} or \gls{vsc}, respectively. 
Under \gls{esc}, it becomes possible to modify the photochemistry/photophysics of molecules including charge transfer processes and electronic spectroscopy, and photoinduced reactions can be influenced~\cite{Wang2014-xn,Kowalewski2016-zo,Triana2018-fy,Eizner2019-ke,Ulusoy2019-uq,Groenhof2019-nz,Felicetti2020-qq,Ulusoy2020-ab,Tichauer2021-mk,Martinez2021-aj,Gudem2022-ej,Couto2022-uv,Li2022-xp,Li2022-dc,Mukherjee2023-mc,Schnappinger23jctc,Weight2023-ma,Bauer2023-gu}. 
Similarly, for \gls{vsc}, the vibrational spectra of molecules are altered by the formation of light-matter hybrid states, and even the chemical reactivity of the ground state can be modified~\cite{George2016-sy,Thomas2016-fy,Thomas2019-ve,Hirai2020-pa,Hirai2020-uv,Ahn2023-qk,Zhong2023-lq,Gu2023-uq,Ahn2023-qk,Weight2023-ma}. 
The observed effects of molecular \gls{esc} and \gls{vsc} are often discussed in a phenomenological way, and understanding of the underlying microscopic and macroscopic physical mechanisms, especially with respect to the effects of \gls{vsc}, is still incomplete~\cite{Schutz2020-en,Simpkins2023-ze,Campos-Gonzalez-Angulo2022-gb,Sidler2022-cg,Davidsson2023-yp,Davidsson2020-bs}. 
In our recent work~\cite{Sidler2023-vm} we have shown numerically that the interaction between an optical cavity and ensembles of molecules not only leads to cavity detuning and a change of the optical length, but also allows for a local molecular polarization mechanism under strong collective vibrational coupling in the thermodynamic limit. 
The interplay of microscopic and macroscopic polarization is due to a cavity-mediated dipole-dipole interaction. A deep understanding of this effect may bridge the gap between existing simplified models for a macroscopic ensemble of molecules and experiments. 
We have been able to study this cavity-mediated dipole-dipole interaction for very large ensembles by using simple Shin-Metiu molecules. As a next step to go beyond this simple molecular model, we present here a formulation of the well-known Hartree-Fock ansatz in the context of the \gls{cboa}~\cite{flick2017atoms,flick2017cavity,Flick2018-ns} derived from the complete non-relativistic Pauli-Fierz Hamiltonian~\cite{spohn2004dynamics,Ruggenthaler2018-ew,jestadt2019light,lindoy2023quantum}. 
We refer to the resulting method as \gls{cbohf} approach and, to the best of our knowledge, this is the first wave function-based method to describe strong coupling of real molecules in a cavity in the \gls{cboa} framework. The \gls{cbohf} method allows us to study cavity-mediated dipole-dipole interactions for realistic molecular systems. 

The first part of this work describes the \gls{cbohf} formalism. Next, the \gls{cbohf} approach is used to explore the effects of collective cavity-mediated coupling in small ensembles of diatomic hydrogen fluoride (\ce{HF}) molecules. 
By explicitly simulating the molecular systems in a self-consistent calculation, we are able to study the interactions within the ensemble beyond what can be captured by scaled single-molecule model Hamiltonians. 
In the following we study static ensembles of molecules but do not address the effects
of vibrational resonances~\cite{li2021cavity,schafer2021shining,lindoy2023quantum}.
Consistent with recently reported results~\cite{sidler2020polaritonic,Sidler2023-vm}, we observe non-negligible cavity-induced energy changes at the local molecular level. By a detailed analysis of these energy changes at the microscopic (single-molecule) and macroscopic (molecular ensemble) level, we can show how the size of the ensemble, the individual molecular orientation, and the change in nuclear configuration can affect these collective interactions.

The physics of a cavity-embedded molecular system is described using the non-relativistic Pauli–Fierz Hamiltonian~\cite{spohn2004dynamics,Ruggenthaler2018-ew,jestadt2019light,lindoy2023quantum}, which is represented in the length gauge,
assuming the dipole approximation and the \gls{cboa}~\cite{flick2017atoms,flick2017cavity,Flick2018-ns}. 
Within the \gls{cboa}, the cavity modes and nuclei are considered to be "slow" degrees of freedom compared to the "fast" electrons and, consequently, only the electrons are treated quantum mechanically. 
In the following bold symbols denote vectors and atomic units ($\hbar=4\pi\varepsilon_0=m_e=1$) are used throughout the paper, unless otherwise indicated. 
For a single-mode cavity, the \gls{cboa} Hamiltonian takes the form
\begin{equation}
\label{eq:h_cbo}
\begin{split}
\hat{H}_{CBO} &= \hat{H}_{el} + \frac{1}{2} \omega_c^2 q_{c}^2 - \omega_c q_{c} \left(\bm{\lambda}_{c} \cdot \bm{\hat{\mu}} \right) + \frac{1}{2} \left(\bm{\lambda}_{c} \cdot \bm{\hat{\mu}} \right)^2\,, 
\end{split}
\end{equation}
where 
\begin{equation}
\begin{split}
\quad \bm{\hat{\mu}} = \bm{\hat{\mu}}_{el} + \bm{\mu}_{Nuc} = -\sum_{i=1}^{N_{el}} \bm{\hat{r}}_i  + \sum_{A=1}^{N_{Nuc}} Z_{A} \bm{R}_A\,,
\end{split}
\end{equation}
represents the molecular dipole operator, which is defined by the operators of the $N_{el}$ electron coordinates $\bm{\hat{r}}$, the classic coordinates $\bm{R}$ of the $N_{Nuc}$ nuclei and the nuclear charge $Z$.
$\hat{H}_{el}$ is the Hamiltonian for the field-free many-electron system, and the second term defines the harmonic potential introduced by the cavity-mediated displacement field, with the photon displacement coordinate $q_{c}$ and $\omega_c $ being the frequency of the cavity mode. 
The third term of Eq.~\ref{eq:h_cbo} describes the dipole coupling between the molecular system and the photon displacement field, which is characterized by the coupling strength $\bm{\lambda}_{c}$. 
The last term is the \gls{dse} operator~\cite{Schafer2020-cb,Sidler2022-cg}, which is an energy contribution that describes the self-polarization of the molecule-cavity system. 
Note that the inclusion of the \gls{dse} contribution is strictly necessary to obtain a finite polarization and a bounded solution \cite{Rokaj2018-ww}. 
Since in practice a finite basis, such as Gaussian basis sets used in most of quantum chemistry methods, also limits the polarization, the lack of a stable ground state is not always observed in numerical calculations.
In the following, we will show that the \gls{dse} term is not only formally needed, but yields inter-molecular interactions in an ensemble of molecules.
The coupling parameter $\bm{\lambda}_{c}$ for a cavity with an effective mode volume $V_{c}$ is defined as follows:
 \begin{equation}
 \label{eq:lam}
\bm{\lambda}_{c} =  \bm{e} \lambda_{c} =  \bm{e}  \sqrt{\frac{4 \pi}{V_{c}}}\,.
\end{equation}
The unit vector $\bm{e}$ denotes the polarization axis of the cavity mode.  In the context of a Fabry-Pérot type cavity, $\bm{\lambda}_{c}$ can be directly related to the electric vacuum field strength $\epsilon_{c}$ via $\bm{\lambda}_{c} = \bm{e} \sqrt{\frac{2}{\omega_c}} \epsilon_{c}$. Without loss of generality, we will use this relation to quantify $\lambda_{c}$ by $\epsilon_c$.

As in the standard Hartree-Fock approach, the \gls{cbohf} electronic wave function of an $N_{el}$-electron molecular system is a Slater determinant of mutually orthonormal spin orbitals $\varphi_i$~\cite{aszabo82-qc}:
\begin{equation}
 \label{eq:sd}
\Psi( \bm{\tau}_1, \bm{\tau}_2, \ldots, \bm{\tau}_N) = \frac{1}{\sqrt{N_{el}!}} \ke{\varphi_1, \varphi_2, \dots \varphi_{i}}\,.
\end{equation} 
Here $\bm{\tau}_N$ is used to denote the complete set of coordinates associated with the $N$-th electron, comprised of the spatial coordinate $\bm{r}_N$ and a spin coordinate. 
Note that the $N_{el}$-electron system described by $\Psi$ can be a single molecule or an ensemble of many molecules. Thus, it is possible to treat cavity-induced interactions and standard Coulomb interactions between molecules in the ensemble in the same way. 
For the special case of the dilute gas limit, that is, the situation in which the electronic structures of different molecules do not overlap and interact, the ensemble Slater determinant may be replaced by a product~\cite{Sidler2023-vm} of individual molecular Slater determinants. Note that the displacement coordinate
$q_c$ of the electric field mode is treated as a parameter in \gls{cbohf} ansatz, analog to the nuclear coordinates, and is thus not part of the wave function. 

Using $\hat{H}_{CBO}$ and $\Psi$, the \gls{cbohf} energy expectation value $E_{CBO}$ can be determined using the standard \gls{scf} procedure \cite{aszabo82-qc}: 
\begin{equation}
\label{eq:e_cbo}
\bigl\langle  E_{CBO} \bigr\rangle = \bigl\langle  \Psi \big| \hat{H}_{el} - \omega_c q_{c} \left(\bm{\lambda}_{c} \cdot \bm{\hat{\mu}} \right) + \frac{1}{2} \left(\bm{\lambda}_{c} \cdot \bm{\hat{\mu}} \right)^2 \big| \Psi  \bigr\rangle + \frac{1}{2} \omega_c^2 q_{c}^2.
\end{equation}
The resulting energy expectation $E_{CBO}$ consists of four energy contributions:
\begin{equation}
\label{eq:e_cbo_part}
E_{CBO}  = E_{el} +  E_{lin} + E_{dse} + E_{dis} \quad  \text{ with} \quad E_{dis} = \frac{1}{2} \omega_c^2 q_{c}^2\,.
\end{equation}
The detailed derivation of all new energy contributions in the \gls{cbohf} ansatz is given in section~S1 of the supporting information. The following discussion of these energies is based on Hartree-Fock matrix elements formulated in the basis of orthonormal spin orbitals.
The first term $E_{el}$ contains all Hartree-Fock energy components of the many-electron system~\cite{aszabo82-qc} and is only indirectly affected by the cavity via the \gls{scf} procedure.
The second term $E_{lin}$ describes the linear part
of the light-matter coupling and is obtained from the dipole coupling between
the photon displacement field, the electrons, and the nuclei.
It can be written as a sum of one-electron integrals formulated in terms of spin orbitals $\varphi_i$ and a parametric nuclear contribution:
\begin{equation}
 \label{eq:linear}
E_{lin} = - \omega_c q_{c} \bigl\langle \Psi \big| \bm{\lambda}_{c} \cdot \bm{\hat{\mu}} \big| \Psi  \bigr\rangle =  - \omega_c q_{c}  \sum_{i=1}^{N_{oc}}  
\bigl\langle \varphi_i \big| \hat{x} \big| \varphi_i \bigr\rangle  - \omega_c q_{c}  \left( \bm{\lambda}_{c} \cdot \bm{\mu}_{Nuc} \right) \  \text{ with} \ \hat{x}=  -\bm{\lambda}_{c} \cdot \bm{\hat{r}}\,.
\end{equation}
The remaining component $E_{dse}$ can be decomposed into a purely electronic term, a mixed electron-nuclear term, and a pure nuclear contribution:
\begin{equation}
 \label{eq:dse}
 \begin{split}
E_{dse} & = \frac{1}{2} \bigl\langle  \Psi \big| \left(\bm{\lambda}_{c} \cdot \bm{\hat{\mu}} \right)^2 \big| \Psi \bigr\rangle  =  E^{
(el)}_{dse} +  E^{
(e\text{-}n)}_{dse}  +  E^{
(nuc)}_{dse}  \\
 &= \frac{1}{2} \bigl\langle  \Psi \big|  \left( \bm{\lambda}_{c}  \cdot \bm{\hat{\mu}}_{el} \right)^2 \big| \Psi \bigr\rangle + \left( \bm{\lambda}_{c}  \cdot \bm{\mu}_{Nuc} \right) \bigl\langle  \Psi \big| \left( \bm{\lambda}_{c}  \cdot \bm{\hat{\mu}}_{el} \right) \big| \Psi \bigr\rangle + \frac{1}{2} \left( \bm{\lambda}_{c}  \cdot \bm{\mu}_{Nuc} \right)^2 \,.
 \end{split}
\end{equation}
To simplify the electronic structure calculations in this work, the classical nuclei are arranged so that their contributions to the total dipole moment are zero ($\bm{\mu}_{Nuc} = 0$, center of charge). 
Thus, the nuclear contribution in Eq.~\ref{eq:linear} as well as $E^{ (e\text{-}n)}_{dse}$ and $E^{
(nuc)}_{dse}$ are zero by definition. More details on the latter two can be found in section S1 of the Supporting Information. The pure electronic contribution $E^{(el)}_{dse}$, contains
a squared electronic position operator $\hat x^2=\hat x_i \hat x_j$, 
which is a two-electron operator but can be decomposed into one-electron contributions  and a two-electron contributions (for details see Eq.~S7 in the supporting information) \cite{Philbin2022-pc,Riso2022-ll,Vu2022-mx,Liebenthal2022-dp}:
\begin{equation}
E^{(el)}_{dse}  =  E^{(1e)}_{dse} + E^{(2e)}_{dse}  = \frac{1}{2} \sum_{i=1}^{N_{oc}}  \bigl\langle \varphi_i \big|  \hat{x}^2 \big| \varphi_i \bigr\rangle +  \frac{1}{2} \left[ \sum_{i,j}^{N_{oc}} \bigl\langle \varphi_i \big| \hat{x} \big| \varphi_i \bigr\rangle  \bigl\langle \varphi_j \big| \hat{x} \big| \varphi_j \bigr\rangle - 
\left| \bigl\langle \varphi_i \big| \hat{x} \big| \varphi_j \bigr\rangle  \right|^2 \right]\,.
\end{equation} 
The one-electron $E^{(1e)}_{dse}$ term retains the quadratic nature of the $\hat x$ operator and behaves like scaled quadrupole tensor elements and describes a localized energy contribution.
The expansion of the two-electron part, $E^{(2e)}_{dse}$ in terms of spin
orbitals, follows a logic similar to that of the derivation of the Coulomb and 
exchange integrals in the regular Hartree-Fock ansatz.
The terms here can be factorized and further decomposed into a dipole-dipole interaction component $E^{(2J)}_{dse}$ and an exchange-like component $E^{(2K)}_{dse}$. 
This exchange-like quantity $E^{(2K)}_{dse}$ vanishes if $\varphi_i$ and $\varphi_j$ have no spatial overlap and therefore describes a localized interaction:
\begin{equation}
\label{eq:de2x}
E^{(2K)}_{dse} =  - \sum_{i,j}^{N_{oc}} \left| \bigl\langle \varphi_i \big| \hat{x} \big| \varphi_j \bigr\rangle  \right|^2
\end{equation} 
The $E^{(2J)}_{dse}$ part can be rewritten as a product of the scaled electronic dipole moment: 
\begin{equation}
\label{eq:de2i}
E^{(2J)}_{dse} = \sum_{i,j}^{N_{oc}} \bigl\langle \varphi_i \big| \hat{x} \big| \varphi_i \bigr\rangle  \bigl\langle \varphi_j \big| \hat{x} \big| \varphi_j \bigr\rangle =  \left( \bm{\lambda}_{c} \cdot \bigl\langle  \bm{\hat{\mu}}_{el} \bigr\rangle \right) \left( \bm{\lambda}_{c} \cdot \bigl\langle  \bm{\hat{\mu}}_{el} \bigr\rangle \right).
\end{equation} 
Unlike $E^{(2K)}_{dse}$, the dipole-dipole interaction $E^{(2J)}_{dse}$ does not require a spatial overlap of the spin orbitals $\varphi_i$ and $\varphi_j$ and thus results in a delocalized interaction. 
These two properties are of special interest when ensembles of molecules are described. 
The last energy contribution $E_{dis}$ in Eq.~\ref{eq:e_cbo_part} is the energy resulting from the photon displacement field~\cite{Schafer2020-cb}.

By replacing the electronic dipole moment $\bm{\hat{\mu}}_{el}$ of the total ensemble with the sum over the individual molecular dipole moments in Eq.~\ref{eq:de2i} we obtain: 
\begin{equation}
\label{eq:de2i_parts}
E^{(2J)}_{dse} = \sum_{m=1}^{N_{mol}} \left[ \left( \bm{\lambda}_{c} \cdot \bigl\langle  \bm{\hat{\mu}}_{el}^{(m)} \bigr\rangle \right) \left(  \bm{\lambda}_{c} \cdot \bigl\langle  \bm{\hat{\mu}}_{el}^{(m)} \bigr\rangle  \right) + \sum_{n \neq m}^{N_{mol}} \left(  \bm{\lambda}_{c} \cdot \bigl\langle  \bm{\hat{\mu}}_{el}^{(m)} \bigr\rangle  \right)\left(  \bm{\lambda}_{c} \cdot \bigl\langle \bm{\hat{\mu}}_{el}^{(n)} \bigr\rangle \right) \right]\,,
\end{equation} 
where the summations run over all molecules $N_{mol}$ in the ensemble. The first term is the local or intra-molecular contribution to $E^{(2J)}_{dse}$ for each individual molecule $m$, while the second product describes the interaction of the molecule $m$ with all other molecules in the ensemble. 
This inter-molecular interaction depends only on the orientation and size of the individual dipole moments, but not on their distance.
This molecular dipole-dipole interaction term $E^{(2J)}_{dse}$ in combination with $E^{(nuc)}_{dse}$ is commonly used as a first-order approximation~\cite{Fischer2021-eq,Fischer2023-kw,Fischer2023-ob} of the \gls{dse} energy. 
Note that the full $E_{dse}$ can also be approximated with the help of permanent dipole moments and transition dipole moments in a resolution of identity approach by summing over excited-electronic states \cite{Gudem2021-um,Couto2022-uv,Weight2023-ma}.

By solving Eq.~\ref{eq:e_cbo} with a \gls{scf} approach, $E_{CBO}$ is minimized for a given configuration of classic nuclei and a fixed photon displacement coordinate (parametric photon field). 
The ground state for the combined electronic-photonic subsystem is obtained by minimizing $E_{CBO}$ with respect to the photon displacement coordinate, which leads to the following expression:
\begin{equation}
\frac{\partial}{\partial q_c} E_{CBO} = \omega_c^2 q_c  - \omega_c  \bigl\langle \Psi \big| \bm{\lambda}_{c} \cdot \bm{\hat{\mu}} \big| \Psi  \bigr\rangle = \omega_c^2 q_c  - \omega_c \left( \bm{\lambda}_{c} \cdot  \bigl\langle \bm{\hat{\mu}} \bigr\rangle \right) \stackrel{\text{!}}{=} 0\,.
\end{equation}
The resulting minimum of $E_{CBO}$ is:
\begin{equation}\label{eq:minEcbo}
q_c =  q_{min} =  \frac{ \left( \bm{\lambda}_{c} \cdot  \bigl\langle \bm{\hat{\mu}} \bigr\rangle \right) }{\omega_c}\,.
\end{equation}
Because we work in the length gauge,
the electric field $\bm{\mathcal{E}}$~\cite{Schafer2020-cb} is:
\begin{align}
	\tfrac{1}{4 \pi}\bm{\mathcal{E}} = \bm{\mathcal{D}} -\bm{\mathcal{P}} = \tfrac{1}{4 \pi} \boldsymbol{\lambda}_{c} \omega_{c} q_{c} - \tfrac{1}{4 \pi}\boldsymbol{\lambda}_{c} \left(\boldsymbol{\lambda}_{c}\cdot \bigl\langle \bm{\hat{\mu}} \bigr\rangle \right) = 0\,,\label{eq:efield}
\end{align}	 
where $\bm{\mathcal{D}}$ is the cavity-mediated displacement field and $\bm{\mathcal{P}}$ is the polarization. Requiring that the transverse electric field vanishes in
the ground state, $\bm{\mathcal{E}}=0$ also leads for Eq.~\ref{eq:minEcbo}.
This demonstrates that minimizing $E_{CBO}$ with respect to $q_c$ is
equivalent to fulfilling the zero transverse electric field condition~\cite{Schafer2020-cb,Sidler2022-cg} in the semi-classical limit, which guarantees a non-radiating ground state. 
From Eq.~\eqref{eq:efield} we also find that $E_{dis} + E_{lin} + E_{dse} = \tfrac{1}{8\pi} \int_{V_c} \hat{\bm{\mathcal{E}}}^2 \mathrm{d} \, {\bm r}$, which is in agreement with Maxwell's equations~\cite{Schafer2020-cb}. Thus, by making the \gls{cboa}, we discard the magnetic contribution to the photonic energy~\cite{Flick2017-jh}

The main aim of this work is to study an ensemble of well-separated molecules interacting with the cavity field. 
We thus assume that the wave functions of different molecules in the ensemble do not overlap and that the Coulomb interaction between them is negligible. 
By satisfying the zero transversal electric field condition (Eq.~\ref{eq:minEcbo}) for such a molecular ensemble, the photon displacement coordinate $q_{min}$ becomes a function of the total ensemble dipole moment.
Thus, $E_{lin}$ and $E_{dis}$ and $E_{dse}$ are not exclusively dependent on local molecular properties, but rather on the total ensemble.

The \gls{cbohf} method has been implemented in the Psi4NumPy environment~\cite{Smith2018-tu}, which is an extension of the PSI4~\cite{Smith2020-kq} electronic structure package. 
All calculations were performed using the aug-cc-pVDZ basis set~\cite{Kendall1992-wu} and the geometry of the isolated single \ce{HF} molecule has been optimized at the Hartree-Fock level of theory. 
Note that we have not re-optimized the geometries of the molecular systems in the cavity; as such, our calculations do not account for any geometry relaxation effects originating from the presence of the cavity. 
In all \gls{cbohf} calculations performed in this work, we consider a single mode, lossless cavity. 
We keep the collective coupling strength $\bm{\lambda}_{c}$ constant by applying a scaling factor of
scaling factor of ${1}/{\sqrt{N_{mol}}}$ to obtain a fixed Rabi splitting for different ensemble sizes and treat $\lambda_0$ as a tunable coupling parameter:
\begin{equation}
\label{eq:coupling}
\bm{\lambda}_{c} = \frac{\lambda_0}{\sqrt{N_{mol}}} \bm{e}
\end{equation} 
Here $\lambda_0$ is equivalent to $\lambda_c$ in Eq.~\ref{eq:lam} in the single molecule case.
As a result, we increase the mode volume $V_c$ of the cavity, but by including more molecules, we keep the average density of molecules $N_{mol}/V_c$ fixed. For $N_{mol} \gg 1$, we would be approaching the thermodynamic limit. Since the main goal of this work is to demonstrate the \gls{cbohf} ansatz and to understand how the different energy contributions to the cavity-mediated interaction under \gls{vsc} behave when increasing the size of the ensemble but keeping $N_{mol}/V_c$ fixed, it is enough to simulate small ensembles. In this work, we restrict ourselves to up to eight \ce{HF} molecules. 
We use an artificially increased coupling strength $\lambda_0$ in the range of 0.004 to 0.04, which corresponds to effective mode volumes (Eq.~\ref{eq:lam}) in the single-molecule case as large as \SI{125.27}{\nano\metre\cubed}~(for $\lambda_0$ 0.004) or as small as \SI{1.25}{\nano\metre\cubed}~(for $\lambda_0$ 0.04). 
To refer to a more intuitive physical quantity, the unscaled coupling strength $\lambda_0$ is quantified in this work by the vacuum electric field strength $\epsilon_c$. The fundamental cavity frequency $\omega_c$ is set to be identical with the first vibrational mode of the uncoupled \ce{HF} molecule, which is  \SI{4467}{\per\centi\meter} for the chosen level of theory. This value of $\omega_c$ is chosen arbitrarily in our static setup, since resonance effects are not likely to play a role in our analysis. All calculations were performed in a reproducible environment using the Nix package manager together with NixOS-QChem \cite{nix} (commit f5dad404) and Nixpkgs (nixpkgs, 22.11, commit 594ef126).

\begin{figure}[htb!]
     \centering
    \includegraphics[width=0.5\textwidth]{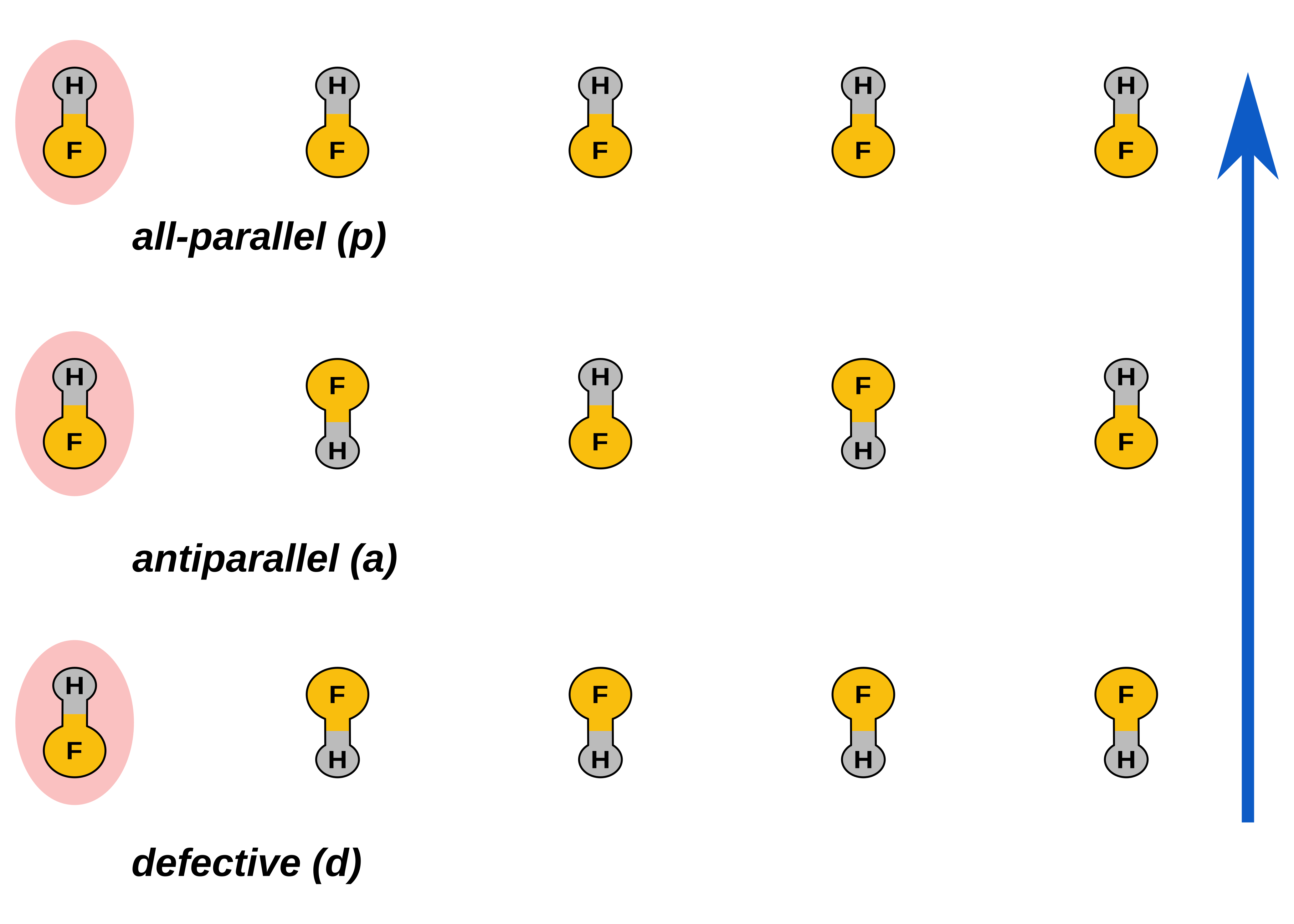}
    \caption{Sketch of the \textit{all-parallel} (p) orientation, the \textit{antiparallel} (a) configuration, and the \textit{defective} (d) orientation shown for an ensemble of five \ce{HF} molecules as an example. All molecules are separated by a distance of \SI{800}{\angstrom}. The scan along the bond length is performed for the \ce{HF} molecule highlighted in red and the cavity polarization axis is shown in blue. Note that except for the \textit{defective} case, the choice of the highlighted molecule is arbitrary.} 
\label{fig:hf_ensembles}
\end{figure}

In this work, we study fixed ensembles of perfectly aligned, well-separated \ce{HF} molecules in an optical cavity. We chose \ce{HF} because of its large permanent dipole moment of \SI{1.8}{\debye}, which is advantageous for the interaction between the molecule and the cavity mode, but its polarizability is small with \SI{0.8}{\angstrom\cubed}.~\cite{GUSSONI1998163} Furthermore, the \ce{HF} simulation results directly extend our previous results in Ref.~\citenum{Sidler2023-vm}, whereas the molecular setup connects to earlier \textit{ab-initio} studies on (collective) electronic strong coupling \cite{sidler2020polaritonic}.
To define these ensembles, the optimized structure of a single \ce{HF} molecule is replicated $N_{mol}$ times. All these replicas are separated by \SI{800}{\angstrom} and placed inside the cavity to avoid interactions via longitudinal electric fields. 
In general, three different orientations of the molecular \ce{HF} ensembles are studied in this work and are visualized in Fig.~\ref{fig:hf_ensembles}. In the first orientation, called \textit{all-parallel}, the \ce{HF} molecules are aligned parallel to the cavity mode polarization axis. In the \textit{antiparallel} case, the $N_{mol}$ \ce{HF} molecules are pairwise antiparallel, resulting in the ensemble dipole moment being zero (even number of molecules) or equal to $\bm{{\mu}}$ of a single \ce{HF} (odd $N_{mol}$). The third configuration of the ensemble, labeled \textit{defective}, represents the situation where the dipole moments of $N_{mol}-1$ \ce{HF} molecules point in the opposite direction to the remaining molecule. Note that for all three ensemble configurations, the individual dipole moment vectors are aligned with the cavity polarization axis, and the zero transverse electric field condition (Eq.~\ref{eq:minEcbo}) is satisfied for the entire ensemble.
This aligned orientation is not the energetically most favorable configuration for the individual molecule (see the supporting information section~S2 for a detailed discussion of the single-molecule situation), but it allows us to set an upper bound on all effects, as it guarantees the maximum molecular cavity interaction. 

\begin{figure}[htb!]
     \centering
         \includegraphics[width=1.0\textwidth]{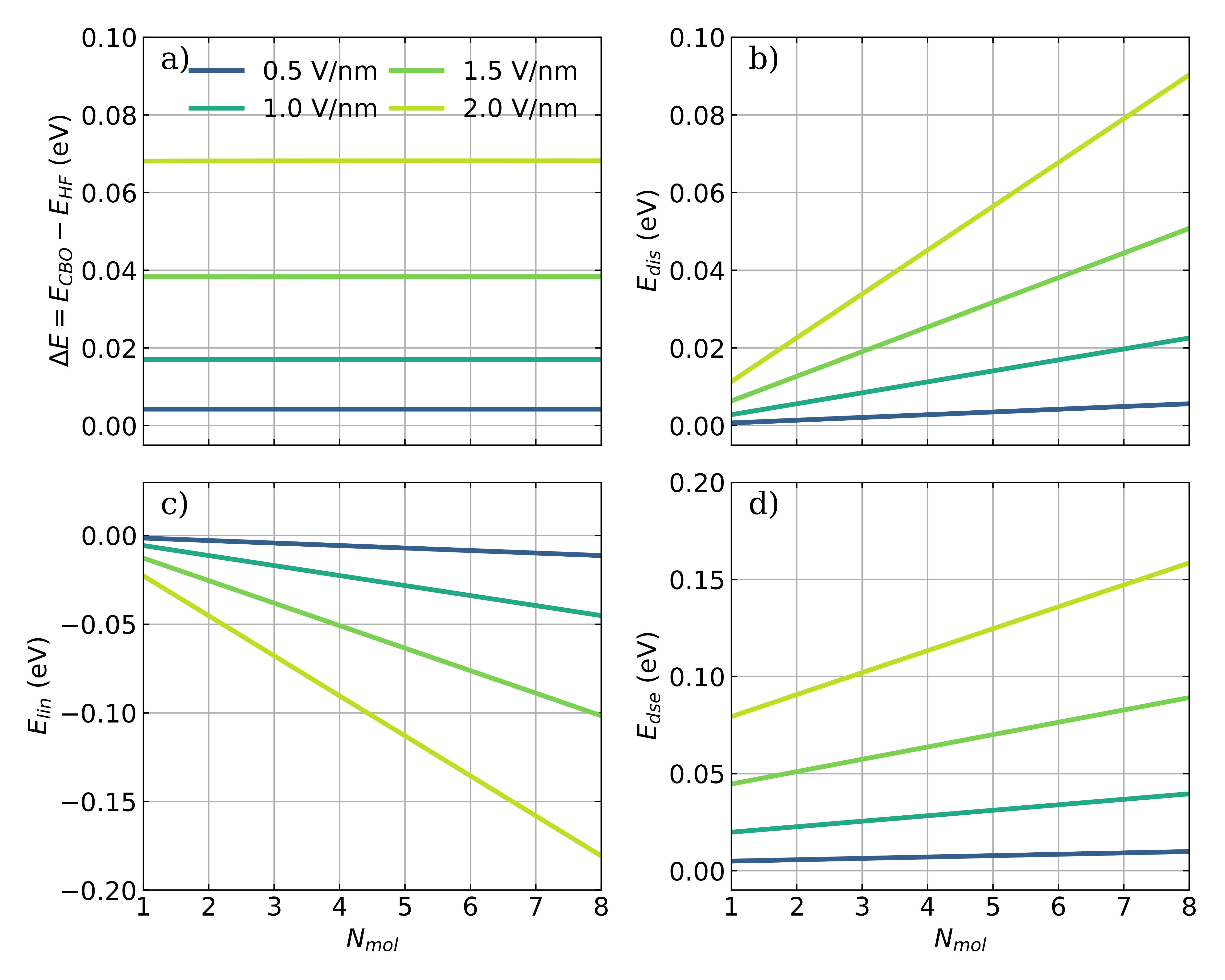}
    \caption{Influence of the cavity interaction on the collective energy of different ensemble sizes and vacuum-field strengths $\epsilon_c$ of \ce{HF} molecules in the \textit{all-parallel} configuration. a) The energy difference $\Delta E$ between $E_{CBO}$ and the field-free energy $E_{HF}$, b) the potential $E_{dis}$ introduced by the cavity-mediated displacement field, c) the contribution of the linear dipole coupling $E_{lin}$, and d) the \gls{dse} contribution $E_{dse}$ for optimized $q_{min}$ as a function of $N_{mol}$. Individual dipole moments are aligned with the cavity polarization axis, and a cavity frequency $\omega_c$ of \SI{4467}{\per\centi\meter} is used. The strength of the cavity field $\epsilon_{c}$ increases from \SI{0.5}{\volt\per\nano\metre} to \SI{2.0}{\volt\per\nano\metre} (color-coded). The coupling strength $\bm{\lambda}_{c}$ used is rescaled according to Eq.~\ref{eq:coupling}.}
\label{fig:hf_ensembel_global}
\end{figure}

All calculations are carried out with rescaled values of $\bm{\lambda}_{c} $, see Eq.~\ref{eq:coupling}. Analogous calculations without rescaling can be found in the supporting information (see Figs.~S5 and~S6 in section~S3). The energy change of the \textit{all-parallel} ensembles induced by the interaction with the cavity, as well as the underlying energy components, is visualized in Fig.~\ref{fig:hf_ensembel_global} as a function of the size of the ensemble $N_{mol}$.

Let us first consider how the different total (ensemble) energies behave and what we can learn from them. For simplicity, we focus on the \textit{all-parallel} configuration. 
In Fig.~\ref{fig:hf_ensembel_global}~a) we see that the proposed rescaling with an increasing ensemble size of the coupling, i.e., the mode volume $V_c \propto N_{mol}$, keeps the light and light-matter interaction energy constant. 
That is, on the total energy level we see that the thermodynamic limiting procedure is well-behaved, and we expect that approximately also for $N_{mol} \gg 1$ we find such an energy difference. From a total energy contribution perspective, one might be tempted to conclude that the photon and photon-matter interaction contribution can be safely ignored, since $E_{el}$ increases linearly with $N_{mol}$ and hence dominates. If we, however, in a next step consider the different contributions of Eq.~\eqref{eq:e_cbo_part}, we see that even for the total-ensemble energy, a delicate balancing of macroscopically scaling energy contributions appears. Indeed, in Fig.~\ref{fig:hf_ensembel_global}~b) we see that the energy of the displacement field increases linearly even if we rescale the coupling strength. This approximately linearly increasing term is countered by $E_{lin}$. That $E_{lin}$ contributes negatively is simple to understand, since the displacement field (interpreted as a constant external field) allows to lower the total energy by separating particles of different charges. Without $E_{dse}$ depicted in Fig.~\ref{fig:hf_ensembel_global}~d), we would find the well-known result that the linear interaction would dissociate and ionize any bound system regardless of the coupling strength~\cite{Rokaj2018-ww, Schafer2020-cb}. We can thus conclude that in order to describe an ensemble of molecules form first principles the dipole self-energy term $E_{dse}$ is needed to find a stable and physical result. 

\begin{figure}[htb!]
     \centering
         \includegraphics[width=1.0\textwidth]{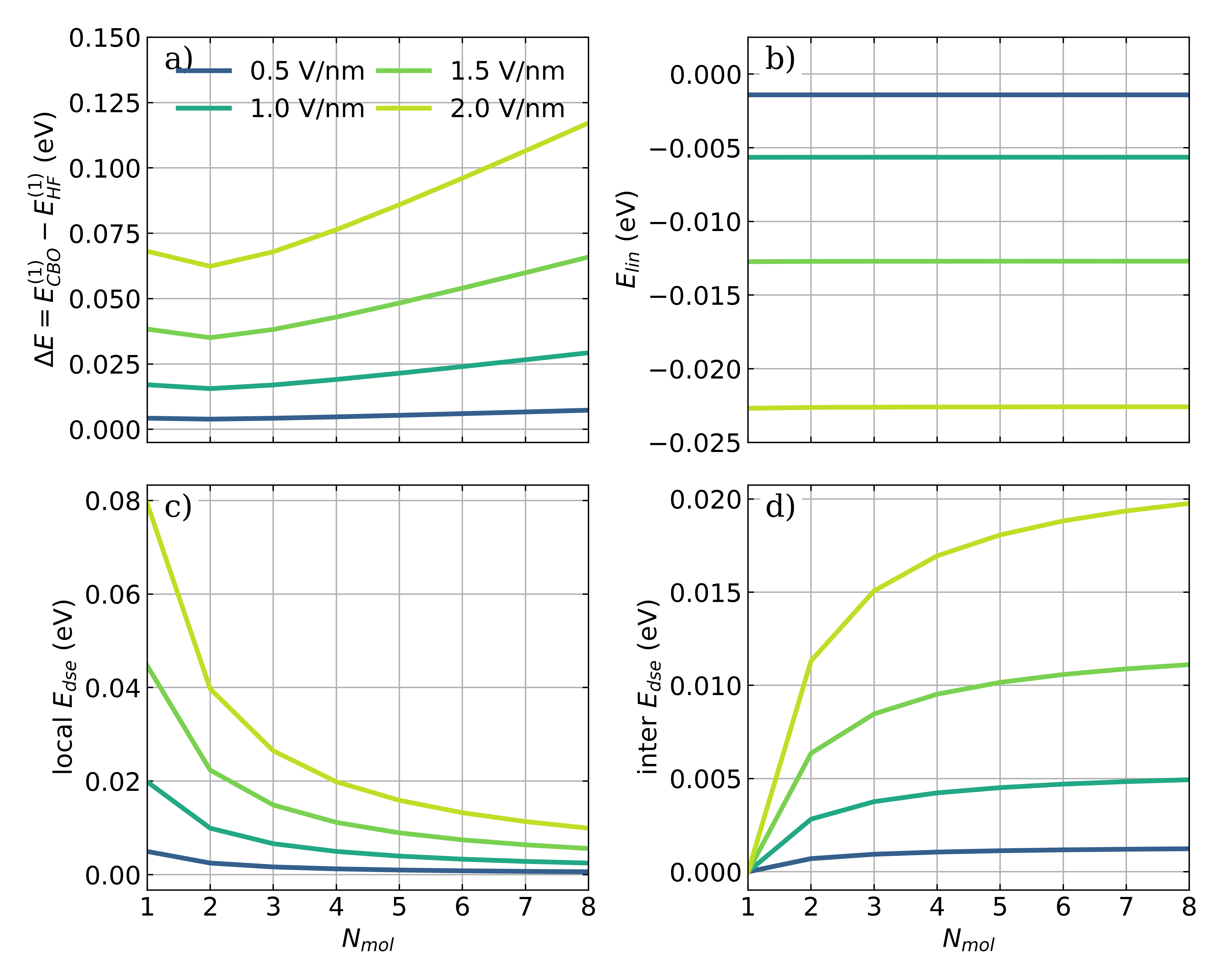}
    \caption{Influence of the cavity interaction on an individual \ce{HF} molecule in \textit{all-parallel} ensembles of different sizes and vacuum field strengths $\epsilon_c$. a) The energy difference $\Delta E$ between $E^{(1)}_{CBO}$ and the field-free energy $E^{(1)}_{HF}$, b) the local linear coupling contribution $E_{lin}$, c) the \textit{local} $E_{dse}$ and d) the \textit{inter} $E_{dse}$ that is part of $E^{(2J)}_{dse}$  (for definition, see Eq.~\ref{eq:de2i_parts}) as a function of $N_{mol}$. The individual dipole moments are aligned with the cavity polarization axis, and a cavity frequency $\omega_c$ of \SI{4467}{\per\centi\meter} is used. The strength of the cavity field $\epsilon_{c}$ increases from \SI{0.5}{\volt\per\nano\metre} to \SI{2.0}{\volt\per\nano\metre} (color-coded). The coupling strength $\bm{\lambda}_{c}$ used is rescaled according to Eq.~\ref{eq:coupling}.} 
\label{fig:hf_ensembel_lcoal_v2}
\end{figure}

So far, we have discussed the effect of the collective coupling on the total molecular ensemble. However, the main question in polaritonic chemistry is to understand how a collectively coupled ensemble can influence individual molecules. In the next step, we will thus analyze the energy changes at the level of a single molecule, which arise as a result of the collective interactions of the entire ensemble.
Such a local perspective is possible since the ensemble \gls{cbohf} density matrix is block diagonal, and individual blocks can be used to create partial density matrices for each molecular subsystem. 
These partial density matrices can be combined with the ensemble Hamiltonian to calculate local energies (per molecule) and combine them pairwise to calculate the interaction between the molecules (see Eq.~\ref{eq:de2i_parts}). This part of the \gls{dse} we denote \textit{inter} $E_{dse}$, while all other contributions to the \gls{dse} are summed and labeled \textit{local} $E_{dse}$. The individual energy obtained per molecule is equivalent to the eigenvalues of the cavity-Hartree equation in our previous work~\cite{Sidler2023-vm}. The change in individual molecular energy induced by the interaction with the cavity, as well as the underlying energy components, is visualized in Fig.~\ref{fig:hf_ensembel_lcoal_v2} as a function of the size of the \textit{all-parallel} ensemble $N_{mol}$.

The energy difference $\Delta E$ between the individual molecular energy $E^{(1)}_{CBO}$ and $E^{(1)}_{HF}$ of the isolated \ce{HF} molecule without cavity interaction is shown in Fig.~\ref{fig:hf_ensembel_lcoal_v2}~a). Note that $E_{dis}$ is an ensemble quantity, but since each molecule in the ensemble is affected by the same potential, we include it in $E^{(1)}_{CBO}$. 
If a second molecule is included in the cavity, the local $E^{(1)}_{CBO}$ decreases. As more and more \ce{HF} molecules are added, the initial trend reverses and $E^{(1)}_{CBO}$ increases almost linearly, following the linear behavior of $E_{dis}$ shown in Fig.~\ref{fig:hf_ensembel_global}~b). 
Without $E_{dis}$, $E^{(1)}_{CBO}$ converges to a finite non-zero value with $N_{mol}$ increasing, as can be seen in Fig.~S4~a) in the supporting information.
The local dipole cavity interaction $E_{lin}$ converges to a constant value with increasing $N_{mol}$, as shown in Fig.~\ref{fig:hf_ensembel_lcoal_v2}~b).
This behavior is a direct consequence of fulfilling the zero-field condition for the entire \textit{all-parallel} ensemble, and cannot be generalized to every nuclear configuration. For this specific orientation, the ensemble dipole moment $\bm{\mu}$ increases linearly with the number of molecules, and thus the displacement induced by the cavity leads to higher values of $q_{min}$. 
This effect, in combination with the rescaled coupling$\bm{\lambda}_{c}$ leads to the constant value of $E_{lin}$ that depends only on the coupling strength. 
On the contrary, the \textit{local} $E_{dse}$, visualized in Fig.~\ref{fig:hf_ensembel_lcoal_v2}~c), decays with $\sfrac{1}{N_{mol}}$ and approaches zero in the large $N_{mol}$ limit. 
The intermolecular dipole-dipole energy ( \textit{inter} $E_{dse}$) shown in Fig.~\ref{fig:hf_ensembel_lcoal_v2}~d) is part of the $E^{(2J)}_{dse}$ term and 
arises as a result of the cavity-mediated interaction of the dipole moment of one molecule with all other molecules in the ensemble
(for the definition, see Eq.~\ref{eq:de2i_parts}). 
This energy contribution increases with an increasing number of molecules in the ensemble and approaches a constant, non-zero value following the behavior of $1-\sfrac{1}{ N_{mol}}$.
All these results are clear indications that the nontrivial interplay of the collective photon displacement effects ($E_{lin}$ in combination with $E_{dis}$) and the cavity-mediated dipole-dipole interaction allow for local strong coupling to emerge.

\begin{figure}[htb!]
     \centering
        \includegraphics[width=0.8\textwidth]{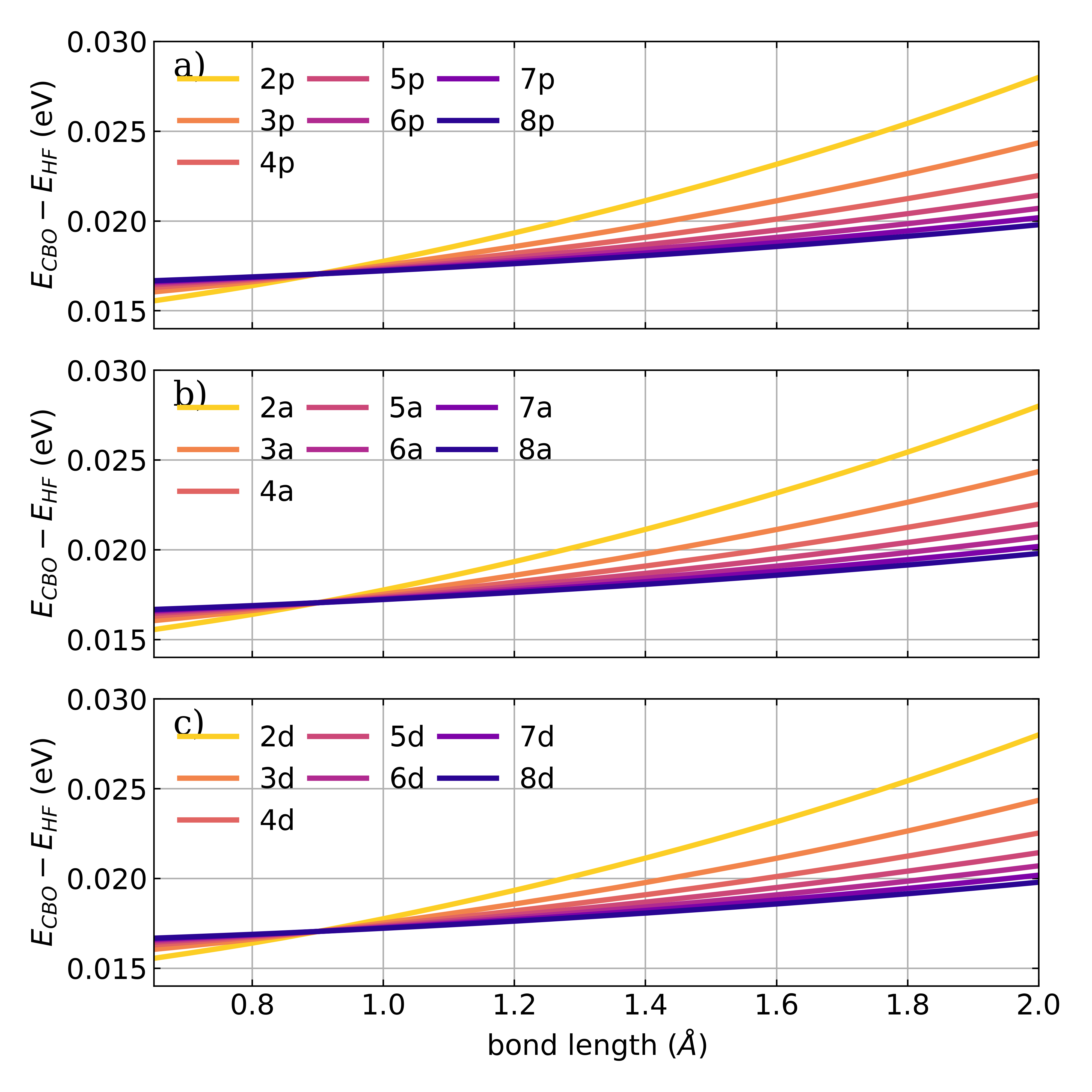}
    \caption{ Cavity-induced energy change along the \ce{HF} bond length for different ensemble sizes and configurations. a) \textit{all-parallel} configuration, b) \textit{antiparallel} configuration, and c) \textit{defective} configuration, for definitions, see Fig.~\ref{fig:hf_ensembles}. A cavity frequency $\omega_c$ of \SI{4467}{\per\centi\meter} is used and the strength of the cavity field $\epsilon_{c}$ is set to \SI{1.5}{\volt\per\nano\metre}. The coupling strength $\bm{\lambda}_{c}$ used is rescaled according to Eq.~\ref{eq:coupling} and the number of molecules in the ensemble is color-coded.}
\label{fig:hf_scan_full}
\end{figure}

In the following, we introduce a defect in perfectly aligned ensembles to further study the effect of anisotropic ensembles on a single molecule.
We perform scans along the bond length of one \ce{HF} molecule in fixed ensembles of different sizes for all three configurations \textit{all-parallel}, \textit{antiparallel} and \textit{defective}. The scan along the bond length is performed for the \ce{HF} molecule encircled in red in Fig.~\ref{fig:hf_ensembles}. In the \textit{defective} orientation, the fixed $N_{mol}-1$ \ce{HF} molecules point in the opposite direction to the perturbed molecule. 
For the resulting two-dimensional \glspl{cpes} spanned by the bond length coordinate and the photon displacement coordinate, the minimum energy path along the bond length is determined. This is equivalent to satisfying the zero transverse electric field condition (Eq.~\ref{eq:minEcbo}) for each nuclear configuration. The energy differences between these minimum energy paths and the field-free one-dimensional \gls{pes} for all three orientations are visualized in Fig.~\ref{fig:hf_scan_full}.
The corresponding energy contributions $E_{lin}$, $E_{dis}$, the \textit{local} $E_{dse}$ and the \textit{inter} $E_{dse}$ are shown in Figs.~S7,~S8, and~S9 in section~S4 of the supporting information.

The observed change on the energetics of the dissociation path due to the cavity interaction is small (see Fig.~\ref{fig:hf_scan_full}), and interestingly, the same for all three ensemble configurations. However, the effect is not negligible and the cavity interaction shifts the ensemble \glspl{cpes} to higher energies compared to the cavity-free \gls{pes}. With increasing bond length, the ensemble dipole moment becomes larger and consequently the upshift mediated by the cavity interaction increases. When comparing the different sizes of the ensemble, there is a decreasing effect on the change in the energy of the ensemble with increasing $N_{mol}$. As a second effect, the energy change is less and less dependent on the bond length. It seems to converge to a non-zero finite value, which is constant with respect to the bond length. A closer look at the individual contributions $E_{lin}$, $E_{dis}$, the \textit{local} $E_{dse}$ and the \textit{inter} $E_{dse}$ all shown in Figs.~S7,~S8, and~S9 in section~S4 of the supporting information can explain these trends. The general shape of the cavity-induced energy change shown in Fig.~\ref{fig:hf_scan_full} is determined by the \textit{local} $E_{dse}$. With an increasing number of molecules in the cavity, this contribution becomes dominated by the fixed ensemble of $N_{mol}-1$ molecules and therefore constant. The other three contributions, $E_{lin}$, $E_{dis}$, and the \textit{inter} $E_{dse}$ are quite large and depend on both the size of the ensemble and the bond length. However, when they are summed, they almost completely cancel each other out, leaving only an almost negligible energy contribution. Since the \textit{local} $E_{dse}$ is the same for all three orientations and dominates the energy change, all three ensemble configurations show the same behavior. It should be noted that by imposing the zero transverse electric field condition along the complete dissociation path, we assume that the whole ensemble coupled to the cavity is in the electronic-photonic ground state. The behavior discussed above may be different if this assumption no longer holds, for example, if the system is coupled to a thermal bath.

\begin{figure}[htb!]
     \centering
         \includegraphics[width=0.6\textwidth]{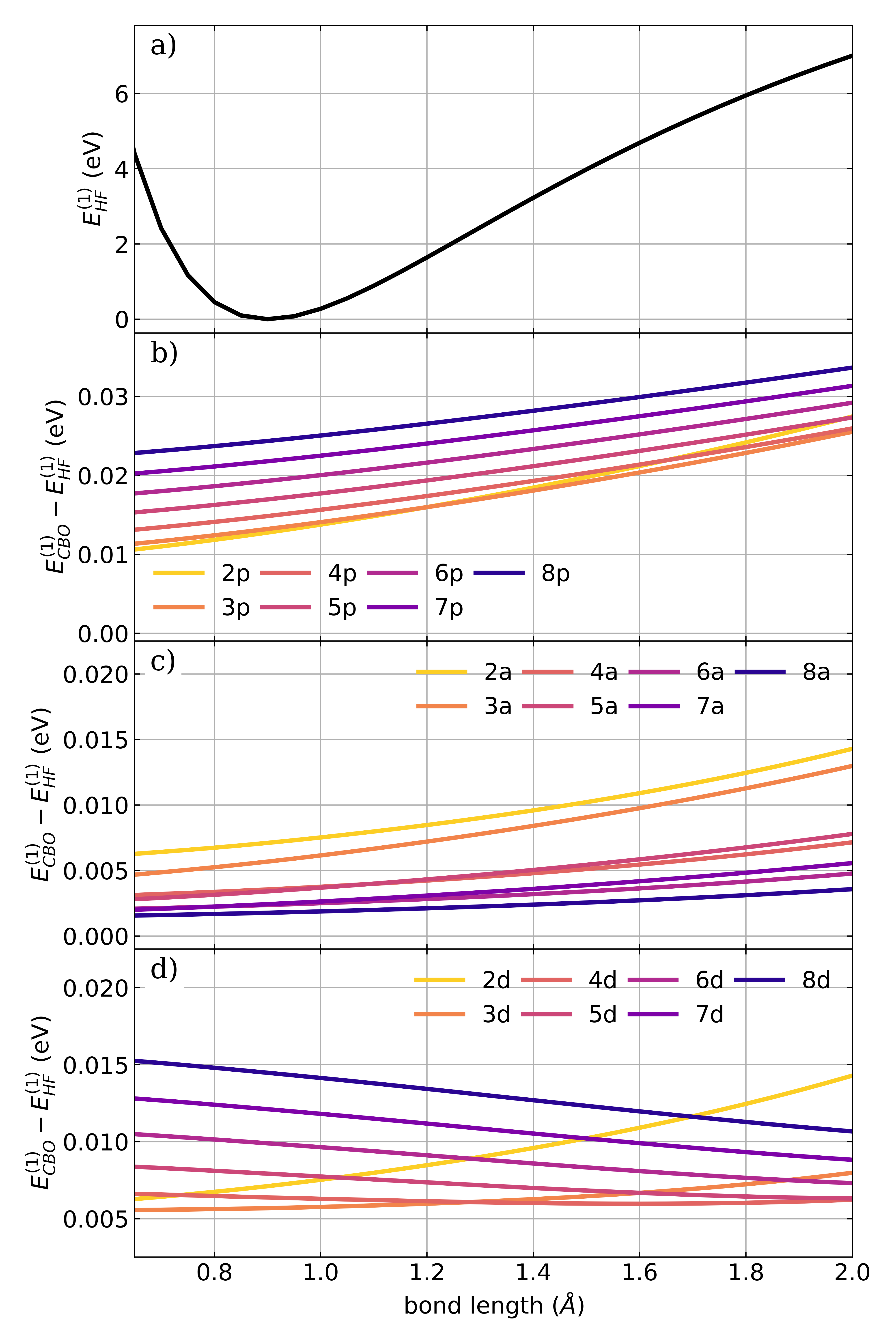}
    \caption{a) Field-free \gls{pes} of a single dissociating \ce{HF} molecule.
    Cavity-induced energy change on the individual dissociating \ce{HF} molecule in ensembles of different size in b) the \textit{all-parallel} configuration, c) \textit{antiparallel} configuration, and d) the \textit{defective} configuration, for definitions see Fig.~\ref{fig:hf_ensembles}. The energy change is calculated as the difference of the locale energy $E_{CBO}^{(1)}$ and the field-free energy $E_{HF}^{(1)}$. A cavity frequency $\omega_c$ of \SI{4467}{\per\centi\meter} is used and the strength of the cavity field $\epsilon_{c}$ is set to \SI{1.5}{\volt\per\nano\metre}. The coupling strength $\bm{\lambda}_{c}$ used is rescaled according to Eq.~\ref{eq:coupling} and the number of molecules in the ensemble is color-coded.} 
\label{fig:hf_scan_local}
\end{figure}

In the last part of this work, we focus on the local perspective of the dissociating \ce{HF} molecule. Its field-free \gls{pes} and the change in the local energy induced by the cavity interaction in the presence of different ensembles are shown in Fig.~\ref{fig:hf_scan_local}.  
Additional figures can be found in section~S4 of the supporting information, showing the underlying energy components: $E_{lin}$ in Fig.~S10, $E_{dis}$ in Fig.~S11, the \textit{local} $E_{dse}$ in Fig.~S12, and the \textit{inter} $E_{dse}$ in Fig.~S13. The last two are calculated from the perspective of the dissociating molecule.

The most striking difference from the perspective of the ensemble is that at the local level of the dissociating \ce{HF} molecule, the three configurations \textit{all-parallel} (Fig.~\ref{fig:hf_scan_local}~b)), \textit{antiparallel} (Fig.~\ref{fig:hf_scan_local}~c)) and \textit{defective} (Fig.~\ref{fig:hf_scan_local}~d)) can be distinguished in cavity-induced energy changes. 
In all three situations, the cavity-induced changes depend on $N_{mol}$ as well as on the length of the bond. 
In the \textit{all-parallel} case, shown in Fig.~\ref{fig:hf_scan_local}~b), the energy change increases with the bond length as well as with the number of molecules. 
Note that the coupling is rescaled by $\sfrac{1}{\sqrt{N_{mol}}}$ to
keep the collective Rabi-splitting constant.
However, the effect on the individual molecules grows with the number of molecules in the ensemble, even though the single molecule coupling strength $\lambda_0$ becomes smaller.
We thus conclude, that collective coupling induces locally strong effects, as has been reported for excited states in \gls{esc} previously~\cite{sidler2020polaritonic}. 
The local contribution $E_{lin}$, which converges to a finite value, is not large enough to compensate for $E_{dis}$, which grows linearly with the size of the ensemble (see Figs.~S10~a) and S11~a)). The resulting upshift in energy is further amplified since \textit{inter} $E_{dse}$ is also positive due to the \textit{all-parallel} configuration and grows with $1-\sfrac{1}{ N_{mol}}$ (see Fig.~S13~a)). Also in the \textit{antiparallel} case (Fig.~\ref{fig:hf_scan_local}~c)) the local energy change due to the cavity interaction increases with increasing bond length. However, the cavity-induced energy change is generally smaller than for the \textit{all-parallel} configuration and becomes significantly smaller with increasing $N_{mol}$. The curves, shown in Fig.~\ref{fig:hf_scan_local}~c), have a pairwise structure, where each ensemble with an even value of $N_{mol}$ is very close in energy to the ensemble with $N_{mol}+1$ molecules. The case of odd and even number of molecules in the \textit{all-parallel} configuration creates two different situations from a single-molecule perspective. For even $N_{mol}$ the whole ensemble reduces to an effective antiparallel bimolecular case, and the situation of odd $N_{mol}$ is equivalent to the single-molecule case where both are scaled down by $\sfrac{1}{N_{mol}}$. Therefore, $E_{lin}$ and $E_{dis}$ become smaller with increasing $N_{mol}$ for odd and even $N_{mol}$ (see Figs.~S10~b) and S11~b)). Furthermore, for even $N_{mol}$ ensembles, there is a small negative contribution from \textit{inter} $E_{dse}$ (see Fig.~S13~b)). In contrast to the \textit{all-parallel} and \textit{antiparallel} configurations, the ensembles in the \textit{defective} orientation show clearly different trends. Similarly to the \textit{all-parallel} case, $\Delta E$ increases as $N_{mol}$ increases for a fixed nuclear configuration, but for $N_{mol} > 3$, 
it simultaneously decreases along the dissociation path. This changing behavior is caused by the interplay of the local $E_{lin}$, the ensemble quantity $E_{dis}$, and the cavity-mediated dipole-dipole interaction. These three contributions are shown in Figs.~S10~c), S11~c), and S12~c). For $N_{mol} < 4$ the ensemble, the dipole moment changes sign along the dissociation path, whereas for a larger ensemble it is negative for the whole pathway. The local dipole moment of the dissociating molecule is in the studied configuration positive, which leads to a positive contribution of the local $E_{lin}$ for $N_{mol} > 3$ and a change in sign for $E_{lin}$ in smaller ensembles. The \textit{inter} $E_{dse}$ (see Fig.~S13~c)) acting on the dissociating molecule is always negative and is increasingly relevant for increasing bond length. Consequently, the cavity-induced energy shift from the local perspective decreases for the \textit{ defective} configuration along the dissociation pathway. In summary, for all three configurations studied, the interaction of the cavity with the molecular ensembles modifies the local energy landscape of the individual molecule. These changes are strongly dependent on the ensemble properties (size and orientation of its components) and are due to the interplay of the displacement field effects and the cavity-induced polarization.

In conclusion, we have established an \textit{ab-initio} Hartree-Fock ansatz in the framework of the cavity Born-Oppenheimer approximation, capable of describing the electronic ground state of molecular systems coupled to an optical cavity. We have applied the \gls{cbohf} ansatz to study the collective effects in small ensembles of diatomic hydrogen fluoride (\ce{HF}) molecules in an optical cavity. 
The detailed analysis of the cavity-induced energy changes for the whole ensemble and individual molecules shows that the self-consistent treatment and the full dipole self-energy operator are crucial to capture relevant aspects for the description of a strongly coupled molecular ensemble and its chemical properties. The \gls{dse} terms are essential to describe cavity-induced polarization at the level of individual molecules, as well as for the whole ensemble. The observed interplay of displacement field effects and the cavity-induced polarization enables energy changes for the individual molecule because of collective coupling to an optical cavity. Consistent with our previous work~\cite{Sidler2023-vm} we could identify a macroscopically induced microscopic polarization mechanism based on intermolecular dipole-dipole interactions. 
Although we have only studied the system in the electronic-photonic ground state, we see indications that thermal fluctuations may play a decisive role in polaritonic chemistry, in line with our previous work~\cite{Sidler2023-vm}. Due to the nature of the intermolecular dipole-dipole interactions, a local change/fluctuation in the dipole moment and/or polarizability could affect the whole ensemble.
A pre-polarization of an ensemble with a static electric field, for example, should lead to an observable effect in experiment. Another interesting topic for further study is the interplay of this self-consistent polarization mechanism with vibrational or electronic resonances. 

The derivation of the \gls{cbohf} equations demonstrates which molecular properties are important for the \gls{dse} term and
the couplings it introduces: molecular dipole moments are important for
inter-molecular interactions, while the combination of dipole moments, quadrupole moments, and transition dipole moments are important on an intra-molecular level. 
The \gls{cbohf} ansatz and the underlying \gls{cboa} formulation offers a suitable framework to derive post-Hartree-Fock methods, such as configuration interaction or coupled cluster, or potential self-consistent embedding schemes~\cite{schaefer2022polaritonic} for molecules under \gls{vsc} or even \gls{esc}~\cite{Schnappinger23jctc}. 
It may also provide potential energy surfaces that can be used
for \textit{ab-initio} semiclassical dynamics or for nuclear-photonic quantum dynamics simulations of molecular ensembles.

\begin{acknowledgement}
This project has received funding from the European Research Council (ERC) under the European Union’s Horizon 2020 research and innovation program (grant agreement no. 852286), the RouTe Project (13N14839), financed by the Federal Ministry of Education and Research (Bundesministerium für Bildung und Forschung (BMBF)) and supported by the European Research Council (ERC-2015-AdG694097), the Cluster of Excellence “CUI: Advanced Imaging of Matter” of the Deutsche Forschungsgemeinschaft (DFG), EXC 2056, project ID 390715994 and the Grupos Consolidados (IT1249-19).
The Flatiron Institute is a division of the Simons Foundation.
\end{acknowledgement}

\begin{suppinfo}
See the supplementary material for the details of the 
derivation of the \gls{cbohf} energy contribution, the discussion of the single-molecule case and additional figures for the ensembles of \ce{HF} molecules in an optical cavity. All data underlying this study are available from the corresponding author upon reasonable request.
\end{suppinfo}

\bibliography{lit.bib}
\providecommand{\latin}[1]{#1}
\makeatletter
\providecommand{\doi}
  {\begingroup\let\do\@makeother\dospecials
  \catcode`\{=1 \catcode`\}=2 \doi@aux}
\providecommand{\doi@aux}[1]{\endgroup\texttt{#1}}
\makeatother
\providecommand*\mcitethebibliography{\thebibliography}
\csname @ifundefined\endcsname{endmcitethebibliography}
  {\let\endmcitethebibliography\endthebibliography}{}

\end{document}


\title{Supporting information: Cavity-Born-Oppenheimer Hartree-Fock Ansatz: \\ Light-matter Properties of Strongly Coupled
Molecular Ensembles}

\author{Thomas Schnappinger}
\affiliation{Department of Physics, Stockholm University, AlbaNova University Center, SE-106 91 Stockholm, Sweden}
  
\author{Dominik Sidler}
\affiliation{Max Planck Institute for the Structure and Dynamics of Matter and Center for Free-Electron Laser Science, Luruper Chaussee 149, 22761 Hamburg, Germany}
\affiliation{The Hamburg Center for Ultrafast Imaging, Luruper Chaussee 149, 22761 Hamburg, Germany}

\author{Michael Ruggenthaler}
\affiliation{Max Planck Institute for the Structure and Dynamics of Matter and Center for Free-Electron Laser Science, Luruper Chaussee 149, 22761 Hamburg, Germany}
\affiliation{The Hamburg Center for Ultrafast Imaging, Luruper Chaussee 149, 22761 Hamburg, Germany}

\author{Angel Rubio}
  \affiliation{Max Planck Institute for the Structure and Dynamics of Matter and Center for Free-Electron Laser Science, Luruper Chaussee 149, 22761 Hamburg, Germany}
    \affiliation{The Hamburg Center for Ultrafast Imaging, Luruper Chaussee 149, 22761 Hamburg, Germany}
  \affiliation{Center for Computational Quantum Physics, Flatiron Institute, 162 5th Avenue, New York, NY 10010, USA}
  \affiliation{Nano-Bio Spectroscopy Group, University of the Basque Country (UPV/EHU), 20018 San Sebasti\'an, Spain}
    
\author{Markus Kowalewski}
\email{markus.kowalewski@fysik.su.se}
\affiliation{Department of Physics, Stockholm University, AlbaNova University Center, SE-106 91 Stockholm, Sweden}

\maketitle

\clearpage
\section{Derivation of the CBO-HF matrix elements}
In this section we provide a detailed derivation of the new matrix elements of the \gls{cbohf} ansatz. The Pauli-Fierz Hamiltonian for a single-mode cavity, in the length gauge and within the dipole approximation, and the \gls{cboa} has the form~\cite{spohn2004dynamics,Ruggenthaler2018-ew,jestadt2019light,lindoy2023quantum,flick2017atoms,flick2017cavity,Flick2018-ns}:
\begin{equation}
\label{eq:h_cbo}
\begin{split}
\hat{H}_{CBO} &= \hat{H}_{el} + \frac{1}{2} \omega_c^2 q_{c}^2 - \omega_c q_{c} \left(\bm{\lambda}_{c} \cdot \bm{\hat{\mu}} \right) + \frac{1}{2} \left(\bm{\lambda}_{c} \cdot \bm{\hat{\mu}} \right)^2\,,
\end{split}
\end{equation}
where 
\begin{equation}
\begin{split}
\quad \bm{\hat{\mu}} = \bm{\hat{\mu}}_{el} + \bm{\mu}_{Nuc} = -\sum_{i=1}^{N_{el}} \bm{\hat{r}}_i  + \sum_{A=1}^{N_{Nuc}} Z_{A} \bm{R}_A\,,
\end{split}
\end{equation}
represents the molecular dipole operator, which is defined by the operators of the $N_{el}$ electron coordinates $\bm{\hat{r}}$, the classic coordinates $\bm{R}$ of the $N_{Nuc}$ nuclei and the nuclear charge $Z$.
Using $\hat{H}_{CBO}$ and a Slater determinant $\Psi$, the \gls{cbohf} energy expectation value $E_{CBO}$ can be determined using the standard \gls{scf} procedure~\cite{aszabo82-qc}: 
\begin{equation}
\label{eq:e_cbo}
\bigl\langle  E_{CBO} \bigr\rangle = \bigl\langle  \Psi \big| \hat{H}_{el} - \omega_c q_{c} \left(\bm{\lambda}_{c} \cdot \bm{\hat{\mu}} \right) + \frac{1}{2} \left(\bm{\lambda}_{c} \cdot \bm{\hat{\mu}} \right)^2 \big| \Psi  \bigr\rangle + \frac{1}{2} \omega_c^2 q_{c}^2
\end{equation}
with the Slater determinant $\Psi$ defined as follows:
\begin{equation}
 \label{eq:sd}
\Psi( \bm{\tau}_1, \bm{\tau}_2, \ldots, \bm{\tau}_N) = \frac{1}{\sqrt{N_{el}!}} \ke{\varphi_1, \varphi_2, \dots \varphi_{i}}.
\end{equation} 
The resulting energy expectation $E_{CBO}$ consists of four energy contributions:
\begin{equation}
\label{eq:e_cbo_part}
E_{CBO}  = E_{el} +  E_{lin} + E_{dse} + E_{dis} \quad  \text{ with} \quad E_{dis} = \frac{1}{2} \omega_c^2 q_{c}^2
\end{equation}
The terms $E_{lin}$ and $E_{dse}$ represent the liner light-matter coupling via the molecular dipole moment and the dipole self energy contribution due to self-polarization, respectively. Both can be extended to one- and two-electron contributions. In the following, we express them in terms of matrix elements of one- and two-electron operators between a $N_{el}$-electron Slater determinant, following the standard rules for Hartree-Fock matrix elements~\cite{aszabo82-qc}. The energy contribution $E_{lin}$ is formulated as modified dipole moment integrals and a parametric nuclear contribution:
\begin{equation}
\begin{split}
E_{lin} &= - \omega_c q_c \bigl\langle \Psi \big| \bm{\lambda}_{c} \cdot \bm{\hat{\mu}_{el}} \big| \Psi  \bigr\rangle   - \omega_c q_c  \left( \bm{\lambda}_{c} \cdot \bm{\mu}_{Nuc} \right) =  \omega_c q_c \sum_{i=1}^{N_{el}} \bigl\langle \Psi \big| \bm{\lambda}_{c} \cdot  \bm{\hat{r}}_i \big| \Psi  \bigr\rangle  - \omega_c q_c  \left( \bm{\lambda}_{c} \cdot \bm{\mu}_{Nuc} \right) \\
&= \omega_c q_c N_{el}  \bigl\langle \Psi \big| \bm{\lambda}_{c} \cdot \bm{\hat{r}} \big| \Psi  \bigr\rangle  - \omega_c q_c  \left( \bm{\lambda}_{c} \cdot \bm{\mu}_{Nuc} \right) =  \omega_c q_{c}  \sum_{i=1}^{N_{oc}}  
\bigl\langle \varphi_i \big| \bm{\lambda}_{c} \cdot \bm{\hat{r}} \big| \varphi_i \bigr\rangle  - \omega_c q_c  \left( \bm{\lambda}_{c} \cdot \bm{\mu}_{Nuc} \right)
\end{split}
\end{equation} 
Before deriving $E_{dse}$ in terms of matrix elements, we decompose the operator describing the \gls{dse} into a purely electronic operator $\hat{H}^{(el)}_{dse}$, a mixed electron-nuclear operator $\hat{H}^{(e\text{-}n)}_{dse}$, and a parametric nuclear energy contribution $E^{(nuc)}_{dse}$. 
\begin{equation}
\begin{split}
\frac{1}{2} \left( \bm{\lambda}_{c} \cdot \bm{\hat{\mu}} \right)^2  =& \frac{1}{2}\left( - \sum_{i=1}^{N_{el}} \bm{\lambda}_{c} \cdot \bm{\hat{r}}_i  +\bm{\lambda}_{c} \cdot \bm{\mu}_{Nuc} \right)^2 \\=& \frac{1}{2}\left( \sum_{i=1}^{N_{el}} \bm{\lambda}_{c} \cdot \bm{\hat{r}}_i   \right)^2 - \left( \sum_{i=1}^{N_{el}} \bm{\lambda}_{c} \cdot \bm{\hat{r}}_i   \right) \left( \bm{\lambda}_{c} \cdot \bm{\mu}_{Nuc} \right) 
+ \frac{1}{2} \left( \bm{\lambda}_{c} \cdot \bm{\mu}_{Nuc} \right)^2 \\=& \frac{1}{2}\left(  \sum_{i=1}^{N_{el}} \bm{\lambda}_{c} \cdot \bm{\hat{r}}_i   \right)^2 - \left( \sum_{i=1}^{N_{el}} \bm{\lambda}_{c} \cdot \bm{\hat{r}}_i   \right) \left( \bm{\lambda}_{c} \cdot \bm{\mu}_{Nuc} \right) 
+  E^{
(nuc)}_{dse} \\
 =& \hat{H}^{(el)}_{dse}  + \hat{H}^{(e\text{-}n)}_{dse} +  E^{(nuc)}_{dse} 
\end{split}
\end{equation}
The purely electronic contribution $E^{(el)}_{dse}$ is decomposed into the one-electron part $E^{(1)}_{dse}$ and the two-electron part $E^{(2)}_{dse}$:
\begin{equation}
\begin{split}
E^{(el)}_{dse}  &= \bigl\langle \Psi \big| \hat{H}^{(el)}_{dse} \big| \Psi \bigr\rangle
 = \frac{1}{2} \bigl\langle \Psi \big| \left( - \sum_{i=1}^{N_{el}} \bm{\lambda}_{c} \cdot \bm{\hat{r}}_i   \right)^2 \big| \Psi \bigr\rangle = \frac{1}{2} \bigl\langle \Psi \big| \left( - \sum_{i=1}^{N_{el}} \bm{\lambda}_{c} \cdot \bm{\hat{r}}_i   \right) \left( - \sum_{j=1}^{N_{el}} \bm{\lambda}_{c} \cdot \bm{\hat{r}}_j   \right) \big| \Psi \bigr\rangle  \\ &= \frac{1}{2} \bigl\langle \Psi \big|  \sum_{i=1}^{N_{el}} \left(\bm{\lambda}_{c} \cdot \bm{\hat{r}}_i \right)^2  +  \sum_{i=1}^{N_{el}} \sum_{j \neq i}^{N_{el}} \left( \bm{\lambda}_{c} \cdot \bm{\hat{r}}_i \right) \left( \bm{\lambda}_{c} \cdot \bm{\hat{r}}_j \right)  \big| \Psi \bigr\rangle \\
&=  \frac{1}{2}\sum_{i=1}^{N_{el}} \bigl\langle \Psi \big| \left(\bm{\lambda}_{c} \cdot \bm{\hat{r}}_i\right)^2  \big| \Psi \bigr\rangle + 
\frac{1}{2} \sum_{i=1}^{N_{el}} \sum_{j \neq i}^{N_{el}} \bigl\langle \Psi \big|
 \left( \bm{\lambda}_{c} \cdot \bm{\hat{r}}_i \right) \left( \bm{\lambda}_{c} \cdot \bm{\hat{r}}_j \right)  \big| \Psi \bigr\rangle = E^{(1e)}_{dse} + E^{(2e)}_{dse} 
\end{split}
\end{equation} 
The energy contribution $E^{(1e)}_{dse}$ can be formulated as modified quadrupole moment integrals:
\begin{equation}
E^{(1e)}_{dse} = \frac{1}{2}\sum_{i=1}^{N_{el}} \bigl\langle \Psi \big| \left( \bm{\lambda}_{c} \cdot \bm{\hat{r}}_i \right)^2  \big| \Psi \bigr\rangle = \frac{1}{2} N_{el}\bigl\langle \Psi \big|  \left( \bm{\lambda}_{c} \cdot \bm{\hat{r}}\right)^2  \big| \Psi 
=  \frac{1}{2} \sum_{i=1}^{N_{oc}}  \bigl\langle \varphi_i \big|  \left( \bm{\lambda}_{c} \cdot \bm{\hat{r}}\right)^2 \big| \varphi_i \bigr\rangle 
\end{equation} 
Since $E^{(2e)}_{dse}$ connects the position operators of two electrons $i$ and $j$, its transformation into Hartree-Fock matrix elements, follows a similar logic as the derivation of the Coulomb interaction in a regular Hartree-Fock ansatz~\cite{aszabo82-qc}:
\begin{equation}
\begin{split}
E^{(2e)}_{dse}  &=\frac{1}{2} \sum_{i=1}^{N_{el}} \sum_{j \neq i}^{N_{el}} \bigl\langle \Psi \big|
 \left( \bm{\lambda}_{c} \cdot \bm{\hat{r}}_i \right) \left( \bm{\lambda}_{c} \cdot \bm{\hat{r}}_j \right)  \big| \Psi \bigr\rangle = \frac{1}{2} N_{el} \left(N_{el}-1\right) \bigl\langle \Psi \big| \left( \bm{\lambda}_{c} \cdot \bm{\hat{r}}_1 \right) \left( \bm{\lambda}_{c} \cdot \bm{\hat{r}}_2 \right)  \big| \Psi \bigr\rangle \\
 &= \frac{1}{2} \sum_{i=1}^{N_{oc}}\sum_{j \neq i}^{N_{oc}}  \bigl\langle \varphi_i \big|  \left( \bm{\lambda}_{c} \cdot \bm{\hat{r}} \right) \big| \varphi_i \bigr\rangle \bigl\langle \varphi_j \big| \left( \bm{\lambda}_{c} \cdot \bm{\hat{r}} \right) \big| \varphi_j \bigr\rangle - \bigl\langle \varphi_i \big|  \left( \bm{\lambda}_{c} \cdot \bm{\hat{r}} \right) \big| \varphi_j \bigr\rangle \bigl\langle \varphi_j \big| \left( \bm{\lambda}_{c} \cdot \bm{\hat{r}} \right) \big| \varphi_i \bigr\rangle 
\end{split}
\end{equation} 
Since the case $i = j$ is equal to zero, the restriction on the summation can be removed:
\begin{equation}
\begin{split}
E^{(2e)}_{dse}  =& \frac{1}{2} \sum_{i=1}^{N_{oc}}\sum_{j=1}^{N_{oc}}  \bigl\langle \varphi_i \big|  \left( \bm{\lambda}_{c} \cdot \bm{\hat{r}} \right) \big| \varphi_i \bigr\rangle \bigl\langle \varphi_j \big| \left( \bm{\lambda}_{c} \cdot \bm{\hat{r}} \right) \big| \varphi_j \bigr\rangle - \bigl\langle \varphi_i \big|  \left( \bm{\lambda}_{c} \cdot \bm{\hat{r}} \right) \big| \varphi_j \bigr\rangle \bigl\langle \varphi_j \big| \left( \bm{\lambda}_{c} \cdot \bm{\hat{r}} \right) \big| \varphi_i \bigr\rangle \\ 
=& \frac{1}{2} \sum_{i=1}^{N_{oc}}\sum_{j=1}^{N_{oc}}  \bigl\langle \varphi_i \big|  \left( \bm{\lambda}_{c} \cdot \bm{\hat{r}} \right) \big| \varphi_i \bigr\rangle \bigl\langle \varphi_j \big| \left( \bm{\lambda}_{c} \cdot \bm{\hat{r}} \right) \big| \varphi_j \bigr\rangle - \big|  \bigl\langle \varphi_i \big|  \left( \bm{\lambda}_{c} \cdot \bm{\hat{r}} \right) \big| \varphi_j \bigr\rangle \big|^2 \\
=&  E^{(2J)}_{dse} +  E^{(2K)}_{dse}
\end{split}
\end{equation} 
The resulting two parts are a Coulomb-like dipole-dipole interaction component $E^{(2J)}_{dse}$ and an exchange-like component $E^{(2K)}_{dse}$, which are calculated via modified dipole moment integrals.
The mixed electron-nuclear operator $\hat{H}^{(e\text{-}n)}_{dse}$ leads to the energy contribution $E^{(e\text{-}n)}_{dse}$, which is formulated as a product of modified dipole moment integrals and a parametric nuclear contribution:
\begin{equation}
\begin{split}
E^{(e\text{-}n)}_{dse}  &=  \bigl\langle \Psi \big| \hat{H}^{(e\text{-}n)}_{dse} \big| \Psi \bigr\rangle  = - \left( \bm{\lambda}_{c} \cdot \bm{\mu}_{Nuc} \right) \sum_{i=1}^{N_{el}}  \bigl\langle \Psi \big|  \bm{\lambda}_{c} \cdot \bm{\hat{r}}_i \big| \Psi \bigr\rangle \\ &=  \left( \bm{\lambda}_{c} \cdot \bm{\mu}_{Nuc} \right) N_{el}\bigl\langle \Psi \big|  \bm{\lambda}_{c} \cdot \bm{\hat{r}} \big| \Psi \bigr\rangle = \left( \bm{\lambda}_{c}  \cdot \bm{\mu}_{Nuc} \right) \sum_{i=1}^{N_{oc}}   \bigl\langle \varphi_i \big| \bm{\lambda}_{c} \cdot \bm{\hat{r}} \big| \varphi_i \bigr\rangle
\end{split}
\end{equation} 
The contribution $E^{(nuc)}_{dse}$ depends only on the nuclear part of the dipole moment and is added as a scalar quantity to $E_{CBO}$.

In the last part of this section we will briefly discuss the underlying modified dipole moment integrals and modified quadrupole moment integrals. For modified dipole moments, the regular integrals are simply multiplied by the corresponding Cartesian
component of the coupling strength $\bm{\lambda}_{c}$ and then summed over the three Cartesian coordinates.  
\begin{equation}
\begin{split}
\bigl\langle \varphi_i \big|  \left( \bm{\lambda}_{c} \cdot \bm{\hat{r}} \right) \big| \varphi_i \bigr\rangle &= \bigl\langle \varphi_i \big|  \lambda_{x}  \hat{r}_{x} + \lambda_{y}  \hat{r}_{y} + \lambda_{z}  \hat{r}_{z} \big| \varphi_i \bigr\rangle \\
 &= \lambda_{x} \bigl\langle \varphi_i \big| \hat{r}_{x} \big| \varphi_i \bigr\rangle + \lambda_{y} \bigl\langle \varphi_i \big|  \hat{r}_{y}  \big| \varphi_i \bigr\rangle + \lambda_{z} 
 \bigl\langle \varphi_i \big| \hat{r}_{z} \big| \varphi_i \bigr\rangle  
\end{split}
\end{equation} 
For the modified quadrupole moment integrals the situation is slightly more complicated. All quadrupole moment tensor elements are multiplied with the two corresponding Cartesian components of the coupling strength $\bm{\lambda}_{c}$ and then summed over all elements.
\begin{equation}
 \begin{alignedat}{3}
& \bigl\langle \varphi_i \big|  \left( \bm{\lambda}_{c} \cdot \bm{\hat{r}}\right)^2 \big| \varphi_i \bigr\rangle  && = && \bigl\langle \varphi_i \big| \left( \lambda_{x}  \hat{r}_{x} + \lambda_{y}  \hat{r}_{y} + \lambda_{z}  \hat{r}_{z} \right)^2 \big| \varphi_i \bigr\rangle \\
& && = &&  \bigl\langle \varphi_i \big|  \lambda_{x}^2 \hat{r}_{x}^2 + \lambda_{y}^2 \hat{r}_{y}^2 + \lambda_{z}^2 \hat{r}_{z}^2 +
2\lambda_{x}\lambda_{y}\hat{r}_{x}\hat{r}_{y} +2\lambda_{x}\lambda_{z}\hat{r}_{x}\hat{r}_{z}
+2\lambda_{y}\lambda_{z}\hat{r}_{y}\hat{r}_{z}
\big| \varphi_i \bigr\rangle \\
& && = && \lambda_{x}^2 \bigl\langle \varphi_i \big|   \hat{r}_{x}^2 \big| \varphi_i \bigr\rangle 
  + \lambda_{y}^2 \bigl\langle \varphi_i \big|   \hat{r}_{y}^2 \big| \varphi_i \bigr\rangle 
  + \lambda_{z}^2 \bigl\langle \varphi_i \big|   \hat{r}_{z}^2 \big| \varphi_i \bigr\rangle +
2\lambda_{x}\lambda_{y}\bigl\langle \varphi_i \big| \hat{r}_{x} \hat{r}_{y} \big| \varphi_i \bigr\rangle \\
& && && +2\lambda_{x}\lambda_{z}\bigl\langle \varphi_i \big| \hat{r}_{x} \hat{r}_{z} \big| \varphi_i \bigr\rangle 
  +2\lambda_{y}\lambda_{z}\bigl\langle \varphi_i \big| \hat{r}_{y} \hat{r}_{z} \big| \varphi_i \bigr\rangle 
\end{alignedat}
\end{equation} 

\clearpage

\section{Single \ce{HF} molecule in an optical cavity}
\begin{figure}[htb!]
     \centering
         \includegraphics[width=0.8\textwidth]{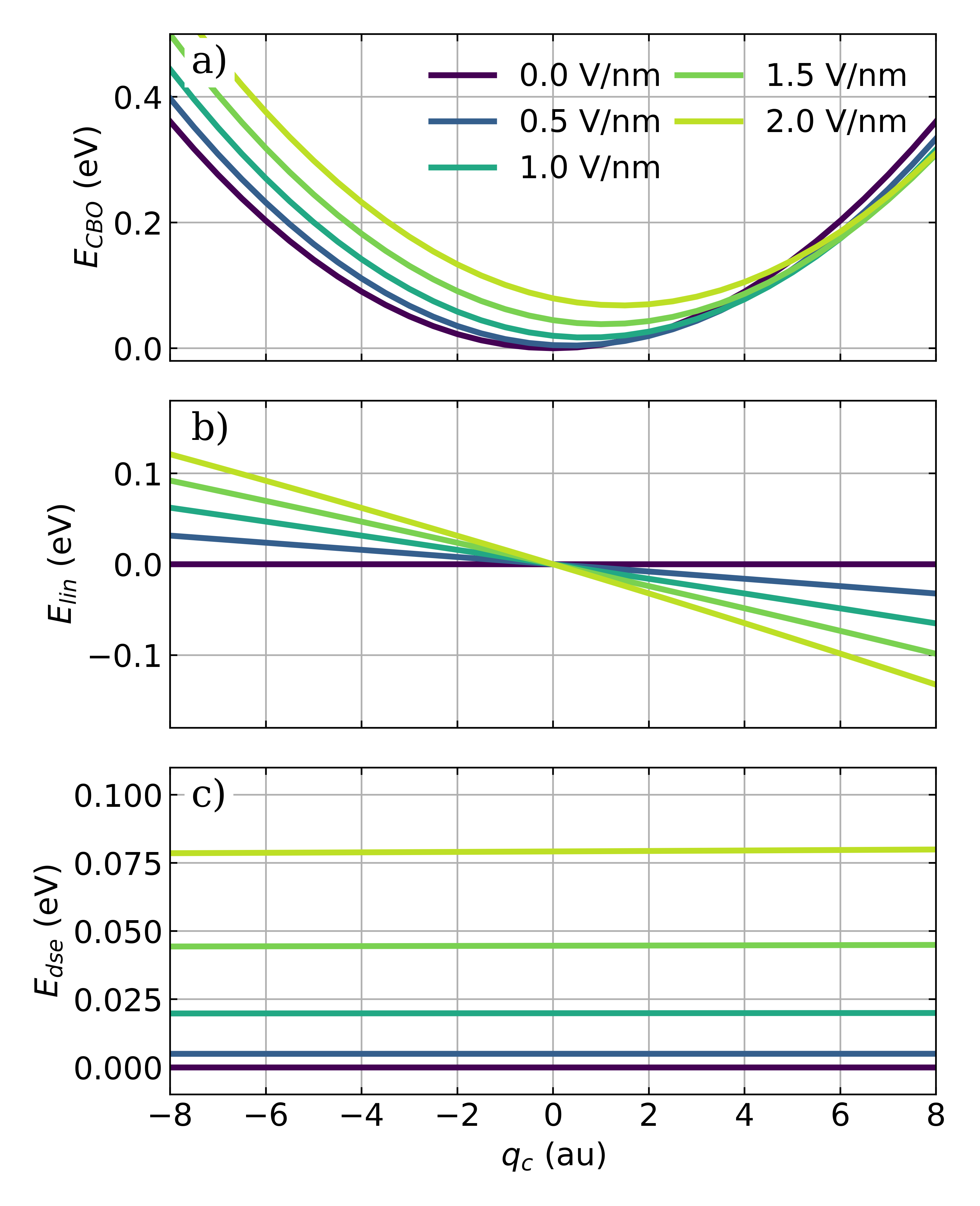}
    \caption{Scan along the photon displacement coordinate $q_c$ for a fixed single \ce{HF} molecule of the total energy $E_{CBO}$  (a)), the linear energy contribution $E_{lin}$ (b)) and the \gls{dse} part $E_{dse}$ (c)). All scans were performed with the molecular dipole moment $\bm{\mu}$ aligned with the cavity polarization axis, a cavity frequency $\omega_c$ of \SI{4467}{\per\centi\meter} and the cavity field strengths $\epsilon_{c}$ is increased from \SI{0.0}{\volt\per\nano\metre} to \SI{2.0}{\volt\per\nano\metre} (color-coded).}
\label{fig:hf_mono}
\end{figure}

The energy change of a single \ce{HF} molecule induced by the interaction with the cavity, as well as the underlying energy components, are visualized in Fig.~\ref{fig:hf_mono}. For this purpose $q_c$ is scanned for different cavity field strengths and a fixed nuclear configuration. The molecular dipole moment $\bm{\mu}$ is aligned with the polarization axis of the cavity.

\begin{figure}[htb!]
     \centering
         \includegraphics[width=1.0\textwidth]{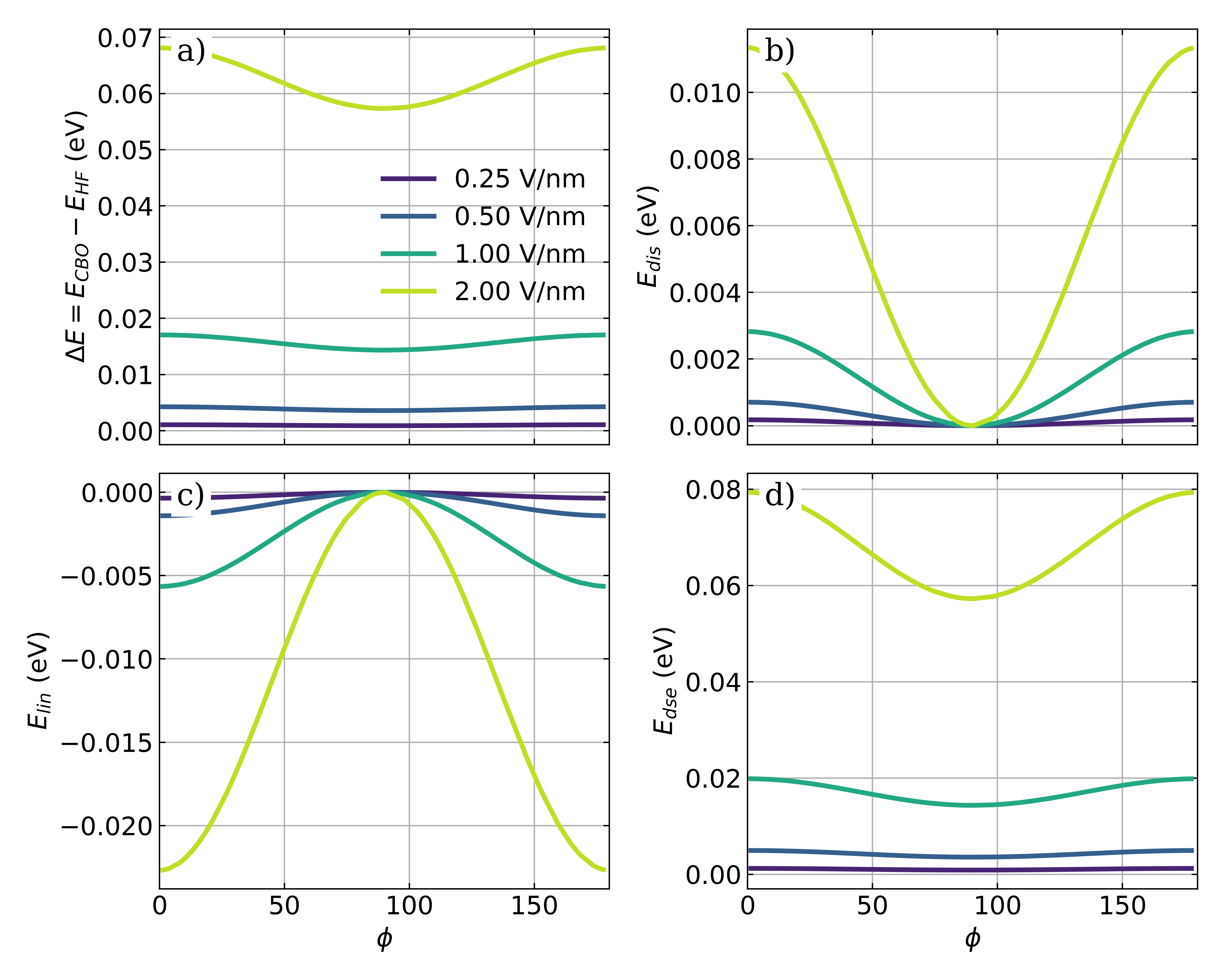}
    \caption{Scan along the angle $\phi$, defined between $\bm{\mu}$ of a single \ce{HF} and the polarization axis $\bm{e}$. a) $\Delta E$ between the rotational \glspl{cpes} and the cavity-free \gls{pes}, b) $E_{dis}$, c) $E_{lin}$, and d) $E_{dse}$ for optimized $q_{min}$. All scans were performed with a cavity frequency $\omega_c$ of \SI{4467}{\per\centi\meter} and the cavity field strengths $\epsilon_{c}$ is increased from \SI{0.25}{\volt\per\nano\metre} to \SI{2.0}{\volt\per\nano\metre} (color-coded).}
\label{fig:hf_mono_rot}
\end{figure}

For the coupling strengths studied, the \glspl{cpes} defined by $E_{CBO}$ (Fig.~\ref{fig:hf_mono}~a)) are basically shifted versions of the harmonic potential $E_{dis}$, for definition, see Eq.~\ref{eq:e_cbo_part}. With increasing $\epsilon_c$ the minimum of the \gls{cpes} defined by $q_{min}$ is shifted to higher values of $q_{c}$ and simultaneously to higher energies. The shift in $q_c$ is due to $E_{lin}$ describing the energy induced by the coupling between the molecule and the photon displacement field. As shown in Fig.~\ref{fig:hf_mono}~b) $E_{lin}$ is a nearly linear function of $q_c$ with a zero crossing at $q_c = 0.0$ and a slope that increases with $\epsilon_c$. The shift of $E_{CBO}$ towards higher energies is caused by $E_{dse}$. This contribution, see  Fig.~\ref{fig:hf_mono}~c), is for the conditions studied nearly constant when changing $q_c$ and its value increases with increasing $\epsilon_c$. To get an impression of how the quantities just discussed ($E_{CBO}$, $E_{lin}$, and $E_{dse}$) depend on the orientation of the molecular dipole moment, the angle $\phi$ between $\bm{\mu}$ and the unit vector pointing along the cavity mode polarization axis is scanned. The energy values obtained for optimized $q_{min}$ are visualized in Fig.~\ref{fig:hf_mono_rot}

\begin{figure}[htb!]
     \centering
         \includegraphics[width=0.8\textwidth]{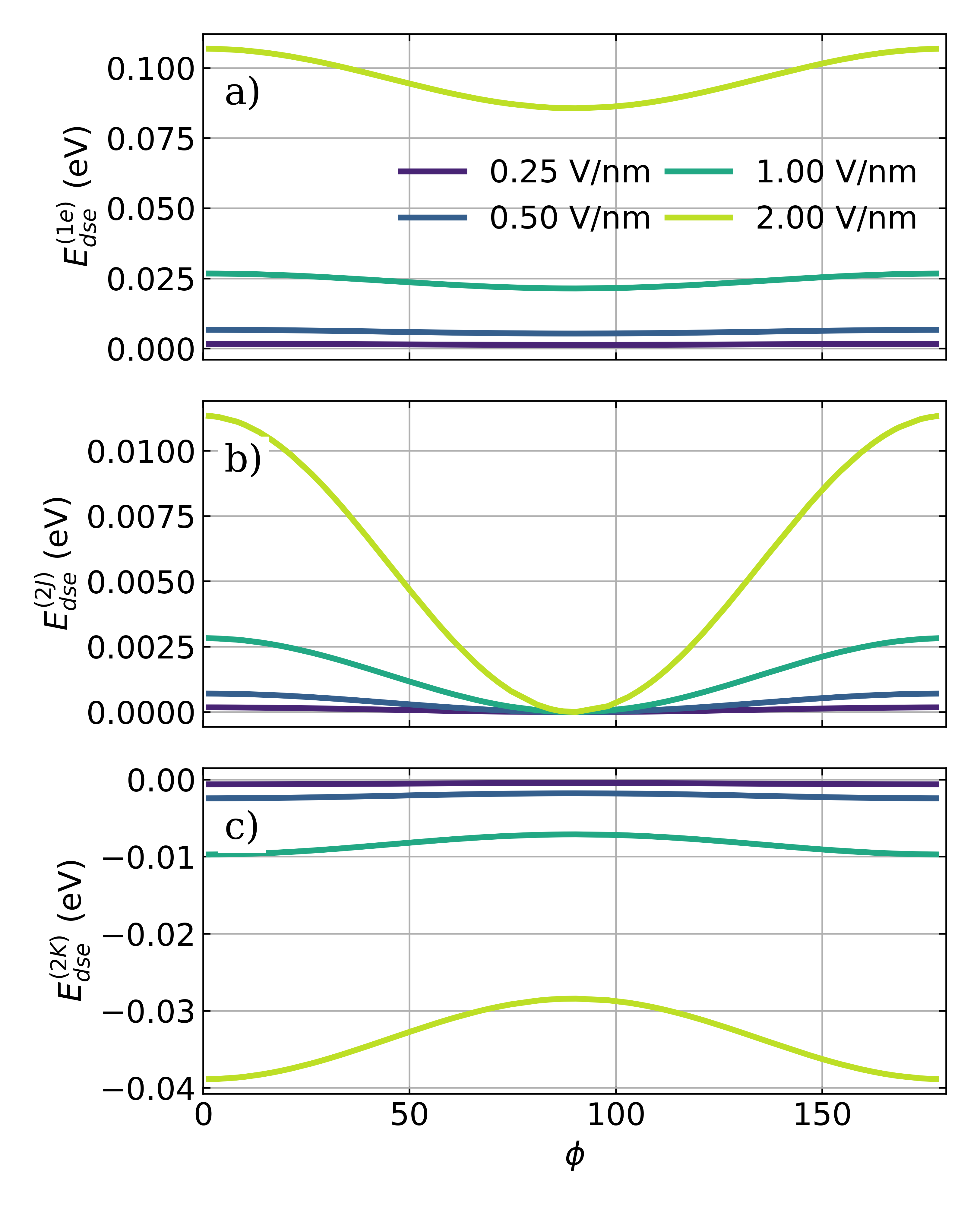}
    \caption{ Scan along the angle $\phi$, defined between $\bm{\mu}$ of a single \ce{HF} and the polarization axis $\bm{e}$. a) $E^{(1e)}_{dse}$ b) $E^{(2J)}_{dse}$ c) $E^{(2K)}_{dse}$. for optimized $q_{min}$. All scans were performed with a cavity frequency $\omega_c$ of \SI{4467}{\per\centi\meter} and the cavity field strengths $\epsilon_{c}$ is increased from \SI{0.25}{\volt\per\nano\metre} to \SI{2.0}{\volt\per\nano\metre} (color-coded).}
\label{fig:hf_mono_dse_rot}
\end{figure}

The \glspl{cpes} shown in Fig.~\ref{fig:hf_mono_rot}~a) have a clear minimum for $\phi = \SI{90}{\degree}$, which corresponds to $\bm{\mu}$ being orthogonal to the cavity polarization axes. Parallel orientation (\SI{0}{\degree}) and antiparallel orientation (\SI{180}{\degree}) are maxima/transition states along the rotation coordinate defined by $\phi$, although the interaction with the cavity is maximal (Fig.~\ref{fig:hf_mono_rot}~b and~c)) for these configurations. For $\phi = \SI{90}{\degree}$ there is no direct dipole cavity interaction, $E_{lin} = \SI{0.0}{\electronvolt}$ see Fig.~\ref{fig:hf_mono_rot}~c) and  $E_{dis} = \SI{0.0}{\electronvolt}$  see Fig.~\ref{fig:hf_mono_rot}~b). However, the energy difference between \glspl{cpes} and cavity-free \gls{pes} is not zero for $\phi = \SI{90}{\degree}$. This is due to $E_{dse}$, which is not zero in this orientation, as shown in Fig.~\ref{fig:hf_mono_rot}~d)). In Fig.~\ref{fig:hf_mono_dse_rot} a further decomposition of $E_{DSE}$ is shown. The overall highest contribution is $E^{(1e)}_{dse}$ and both $E^{(1e)}_{dse}$ and $E^{(2K)}_{dse}$ are non-zero for all possible orientations. Only $E^{(2J)}_{dse}$ is zero for $\phi = \SI{90}{\degree}$. 

\clearpage

\section{Fixed ensembles of \ce{HF} molecules in an optical cavity}

\begin{figure}[htb!]
     \centering
         \includegraphics[width=1.0\textwidth]{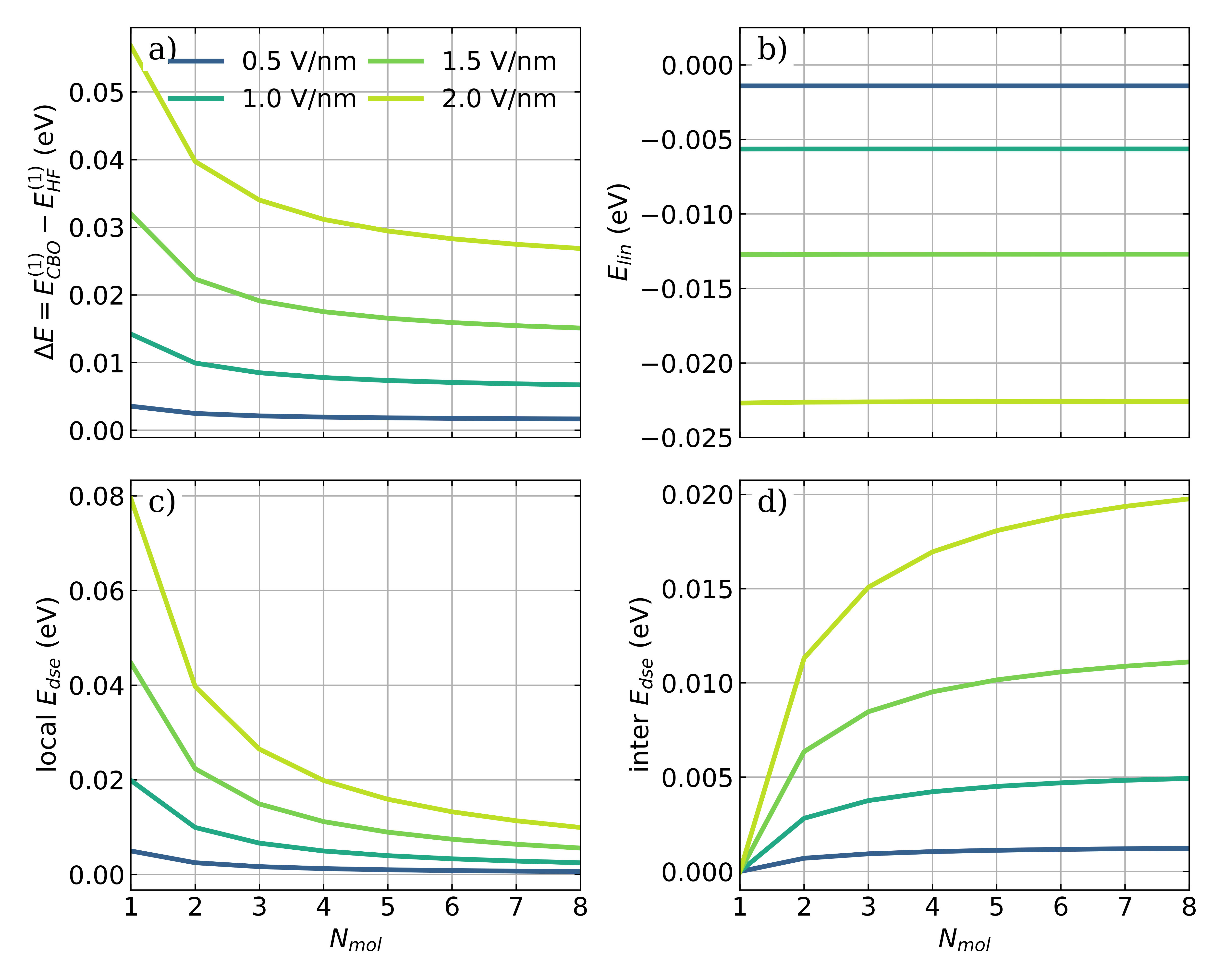}
    \caption{Influence of the cavity interaction on an individual \ce{HF} molecule in \textit{all-parallel} ensembles of different size and vacuum-field strengths $\epsilon_c$. a) The energy difference $\Delta E$ between $E^{(1)}_{CBO}$ and the field-free energy $E^{(1)}_{HF}$ without $E_{dis}$ included, b) the local linear energy contribution $E_{lin}$, c) the local $E_{dse}$ and d) the intermolecular dipole-dipole energy as a function of $N_{mol}$. The individual dipole moments are aligned with the cavity polarization axis, and a cavity frequency $\omega_c$ of \SI{4467}{\per\centi\meter} is used. The strength of the cavity field $\epsilon_{c}$ increases from \SI{0.5}{\volt\per\nano\metre} to \SI{2.0}{\volt\per\nano\metre} (color-coded). The used coupling strength $\bm{\lambda}_{c}$ is rescaled according to Eq.~16.} 
\label{fig:hf_ensembel_lcoal_v2_nD}
\end{figure}

The change in individual molecular energy induced by the interaction with the cavity, as well as the underlying energy components, are visualized in Fig.~\ref{fig:hf_ensembel_lcoal_v2_nD} as a function of the size of the \textit{all-parallel} ensemble $N_{mol}$ without $E_{dis}$ included.

\begin{figure}[htb!]
     \centering
         \includegraphics[width=1.0\textwidth]{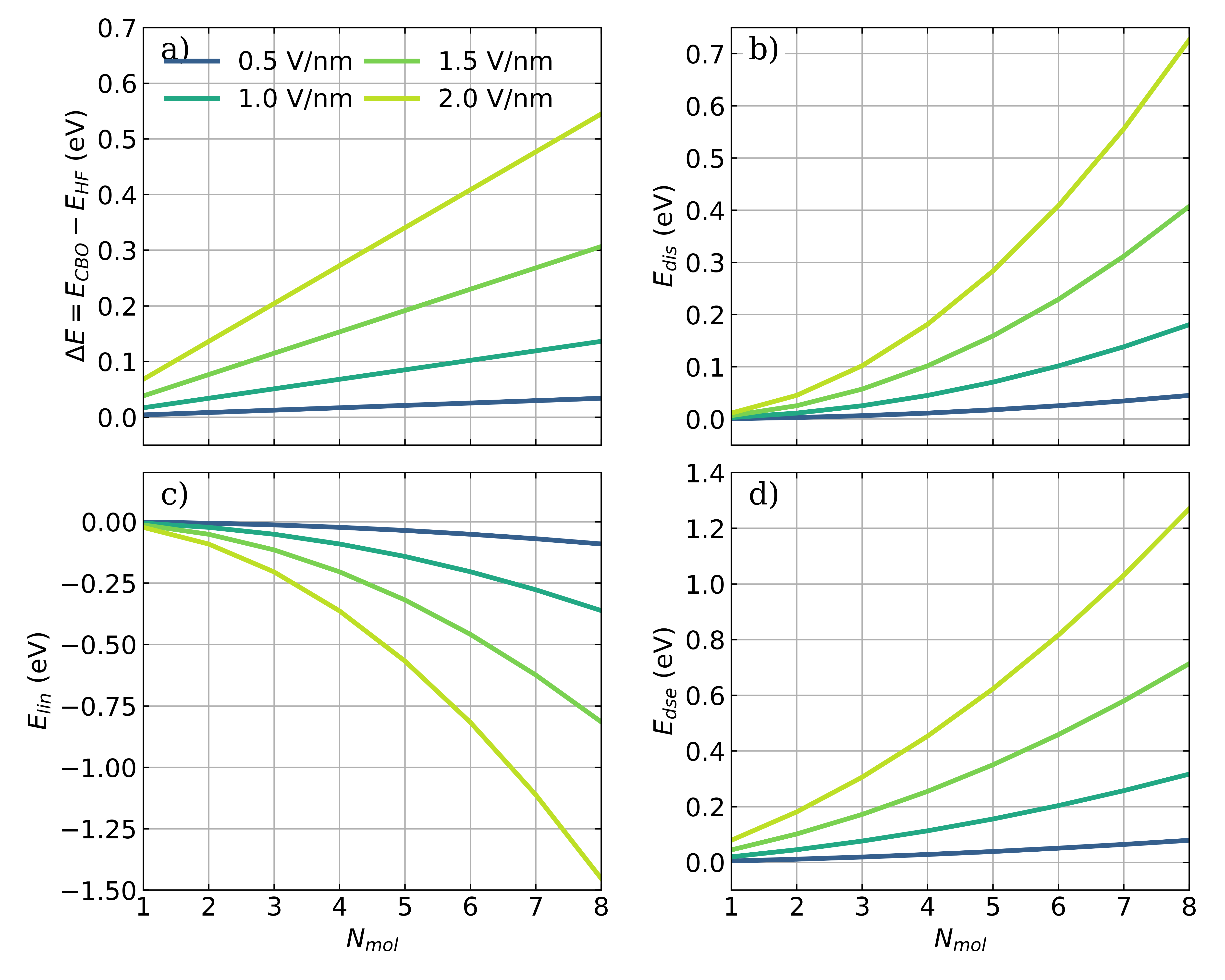}
    \caption{ Influence of the cavity interaction on the collective energy of different ensembles of perfectly aligned \ce{HF} molecules, without rescaling of  $\bm{\lambda}_{c}$. a) The total energy $E_{CBO}$ referenced to the case without a cavity ($E_{HF}$), b) cavity potential $E_{dis}$, c) linear energy contribution $E_{lin}$, and d) the \gls{dse} part $E_{dse}$ for optimized $q_{min}$ as a function of $N_{mol}$. Individual dipole moments are aligned with the cavity polarization axis and a cavity frequency $\omega_c$ of \SI{4467}{\per\centi\meter} is used. The strength of the cavity field $\epsilon_{c}$ increases from \SI{0.5}{\volt\per\nano\metre} to \SI{2.0}{\volt\per\nano\metre} (color-coded).}
\label{fig:hf_ensembel_global_rs}
\end{figure}

\begin{figure}[htb!]
     \centering
         \includegraphics[width=1.0\textwidth]{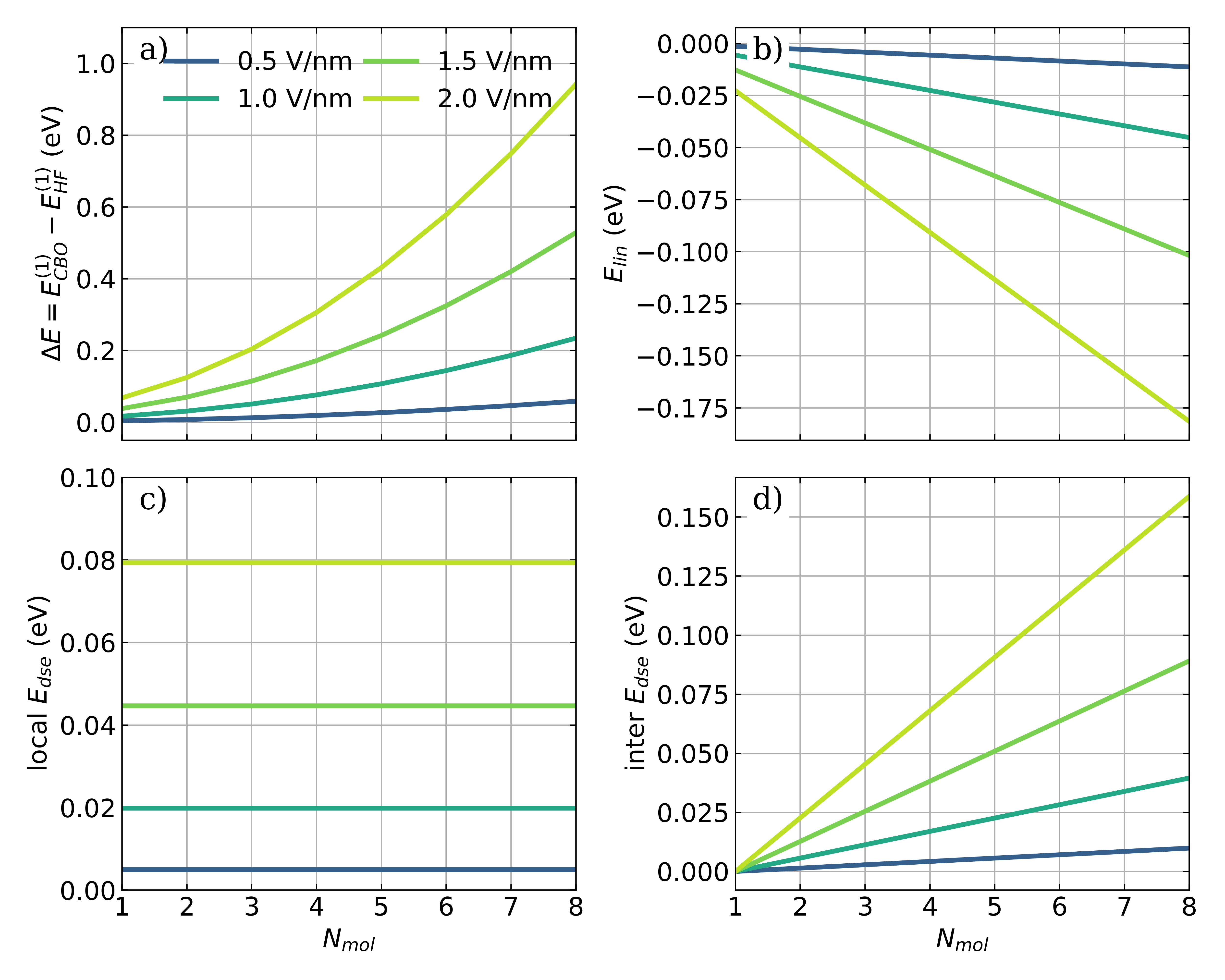}
    \caption{Influence of the cavity interaction on an individual \ce{HF} molecule in ensembles of different size, without rescaling of $\bm{\lambda}_{c}$. a) The individual molecular energy $E_{CBO}$ referenced to the case of a single molecule without cavity interaction, b) the local linear energy contribution $E_{lin}$, c) the local $E_{dse}$ and d) the intermolecular dipole-dipole energy as a function of $N_{mol}$. The individual dipole moments are aligned with the cavity polarization axis and a cavity frequency $\omega_c$ of \SI{4467}{\per\centi\meter} is used. The strength of the cavity field $\epsilon_{c}$ increases from \SI{0.5}{\volt\per\nano\metre} to \SI{2.0}{\volt\per\nano\metre} (color-coded).} 
\label{fig:hf_ensembel_lcoal_v2_rs}
\end{figure}

Supplementary simulation results for fixed ensembles of \ce{HF} molecules in an optical cavity without rescaling $\bm{\lambda}_{c}$ are shown in Figs.~\ref{fig:hf_ensembel_global_rs} and~\ref{fig:hf_ensembel_lcoal_v2_rs}. The results discussed for the scaled case in the manuscript are still valid, and only the scaling behaviors change. For the ensemble perspective, see Fig.~\ref{fig:hf_ensembel_global_rs}, the cavity-induced change in the total energy scales linearly with $N_{mol}$, while its three contributions ($E_{dis}$, $E_{lin}$, and $E_{dse}$) scale quadratic. The energy changes for an individual \ce{HF} molecule, shown in Fig.~\ref{fig:hf_ensembel_lcoal_v2_rs}~a), scales quadratic with $N_{mol}$. The linear interaction $E_{lin}$ (Fig.~\ref{fig:hf_ensembel_lcoal_v2_rs}~b)) and the interacting part of the $E_{dse}$ (Fig.~\ref{fig:hf_ensembel_lcoal_v2_rs}~d)) changing linear with $N_{mol}$ and the local part of $E_{dse}$ (Fig.~\ref{fig:hf_ensembel_lcoal_v2_rs}~c)) is constant as expected.

\clearpage

\section{Scanned ensembles of \ce{HF} molecules in an optical cavity}

\begin{figure}[htb!]
     \centering
         \includegraphics[width=1.0\textwidth]{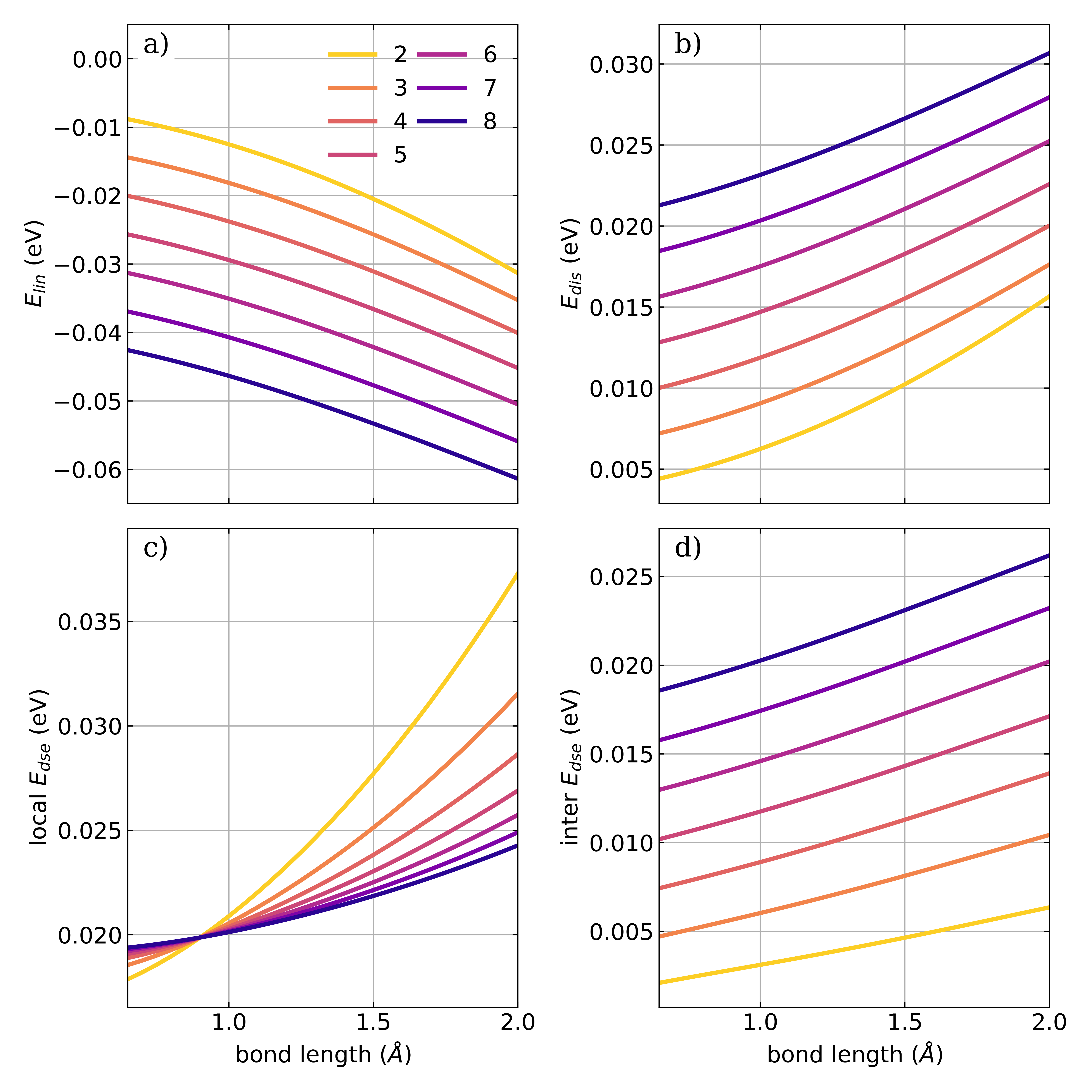}
    \caption{Cavity-induced energy contribution of the complete ensemble along the \ce{HF} bond length for different ensemble sizes in the \textit{all-parallel} configuration. a) linear energy contribution $E_{lin}$, b) cavity potential $E_{dis}$, c) local part of $E_{dse}$ d) interaction part of $E_{dse}$. A cavity frequency $\omega_c$ of \SI{4467}{\per\centi\meter} is used. The strength of the cavity field $\epsilon_{c}$ is \SI{1.5}{\volt\per\nano\metre} and the number of molecules in the ensemble is color-coded.The used coupling strength $\bm{\lambda}_{c}$ is rescaled according to Eq.~16.} 
\label{fig:hf_scan_part_para_ens}
\end{figure}

\begin{figure}[htb!]
     \centering
         \includegraphics[width=1.0\textwidth]{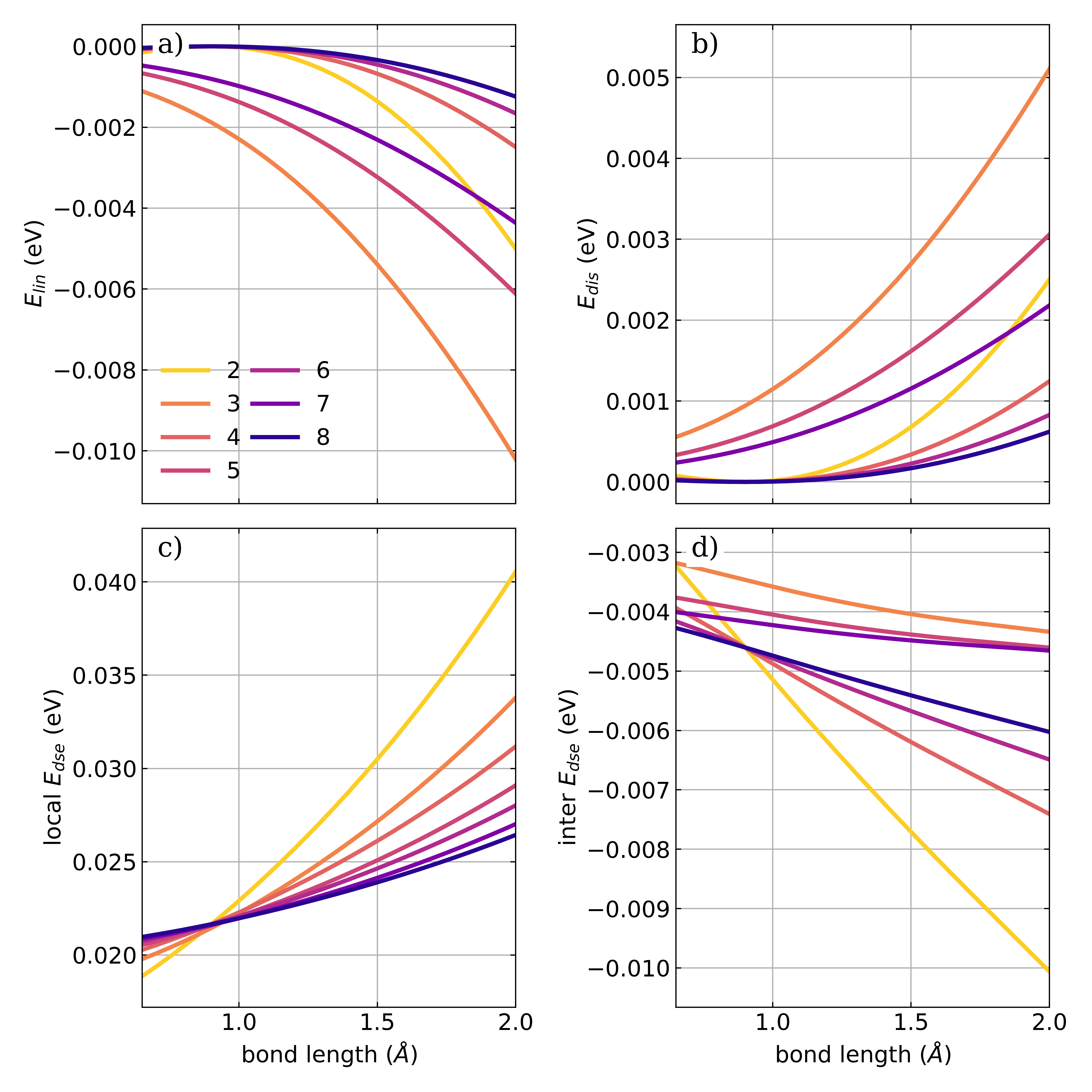}
    \caption{Cavity-induced energy contribution of the complete ensemble along the \ce{HF} bond length for different ensemble sizes in the \textit{antiparallel} configuration. a) linear energy contribution $E_{lin}$, b) cavity potential $E_{dis}$, c) local part of $E_{dse}$ d) interaction part of $E_{dse}$. A cavity frequency $\omega_c$ of \SI{4467}{\per\centi\meter} is used. The strength of the cavity field $\epsilon_{c}$ is \SI{1.5}{\volt\per\nano\metre} and the number of molecules in the ensemble is color-coded.The used coupling strength $\bm{\lambda}_{c}$ is rescaled according to Eq.~16.} 
\label{fig:hf_scan_part_anti_ens}
\end{figure}

\begin{figure}[htb!]
     \centering
         \includegraphics[width=1.0\textwidth]{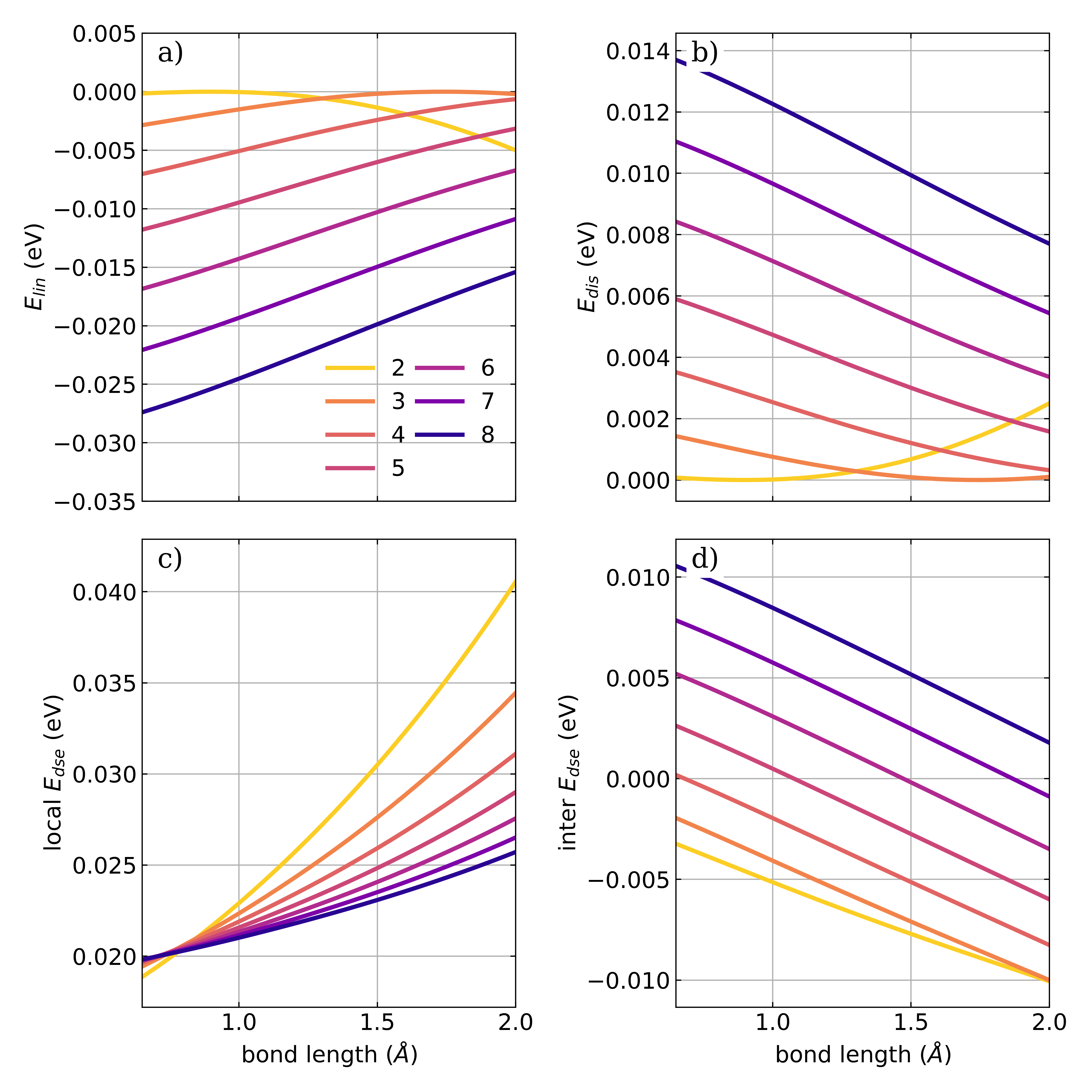}
    \caption{Cavity-induced energy contribution of the complete ensemble along the \ce{HF} bond length for different ensemble sizes in the \textit{defective} configuration. a) linear energy contribution $E_{lin}$, b) cavity potential $E_{dis}$, c) local part of $E_{dse}$ d) interaction part of $E_{dse}$. A cavity frequency $\omega_c$ of \SI{4467}{\per\centi\meter} is used. The strength of the cavity field $\epsilon_{c}$ is \SI{1.5}{\volt\per\nano\metre} and the number of molecules in the ensemble is color-coded.The used coupling strength $\bm{\lambda}_{c}$ is rescaled according to Eq.~16.} 
\label{fig:hf_scan_part_dis_ens}
\end{figure}

\begin{figure}[htb!]
     \centering
         \includegraphics[width=0.8\textwidth]{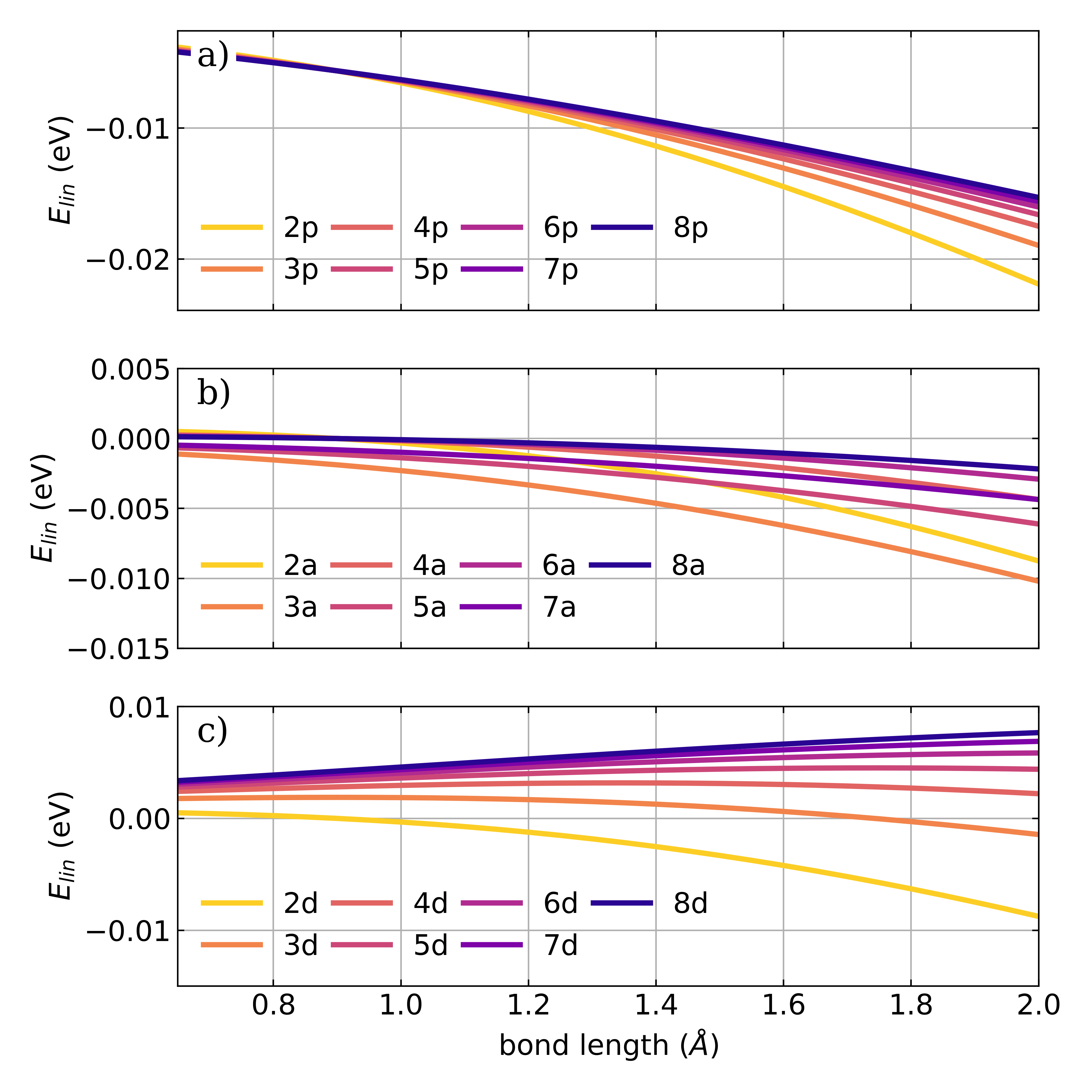}
    \caption{The local linear energy contribution $E_{lin}$ along the \ce{HF} bond length for different ensemble sizes is shown for a) the \textit{all-parallel} configuration, b) the \textit{antiparallel} configuration and c) the \textit{defective} configuration. A cavity frequency $\omega_c$ of \SI{4467}{\per\centi\meter} is used. The strength of the cavity field $\epsilon_{c}$ is \SI{1.5}{\volt\per\nano\metre} and the number of molecules in the ensemble is color-coded.The used coupling strength $\bm{\lambda}_{c}$ is rescaled according to Eq.~16.}
\label{fig:hf_scan_linear}
\end{figure}

\begin{figure}[htb!]
     \centering
         \includegraphics[width=0.8\textwidth]{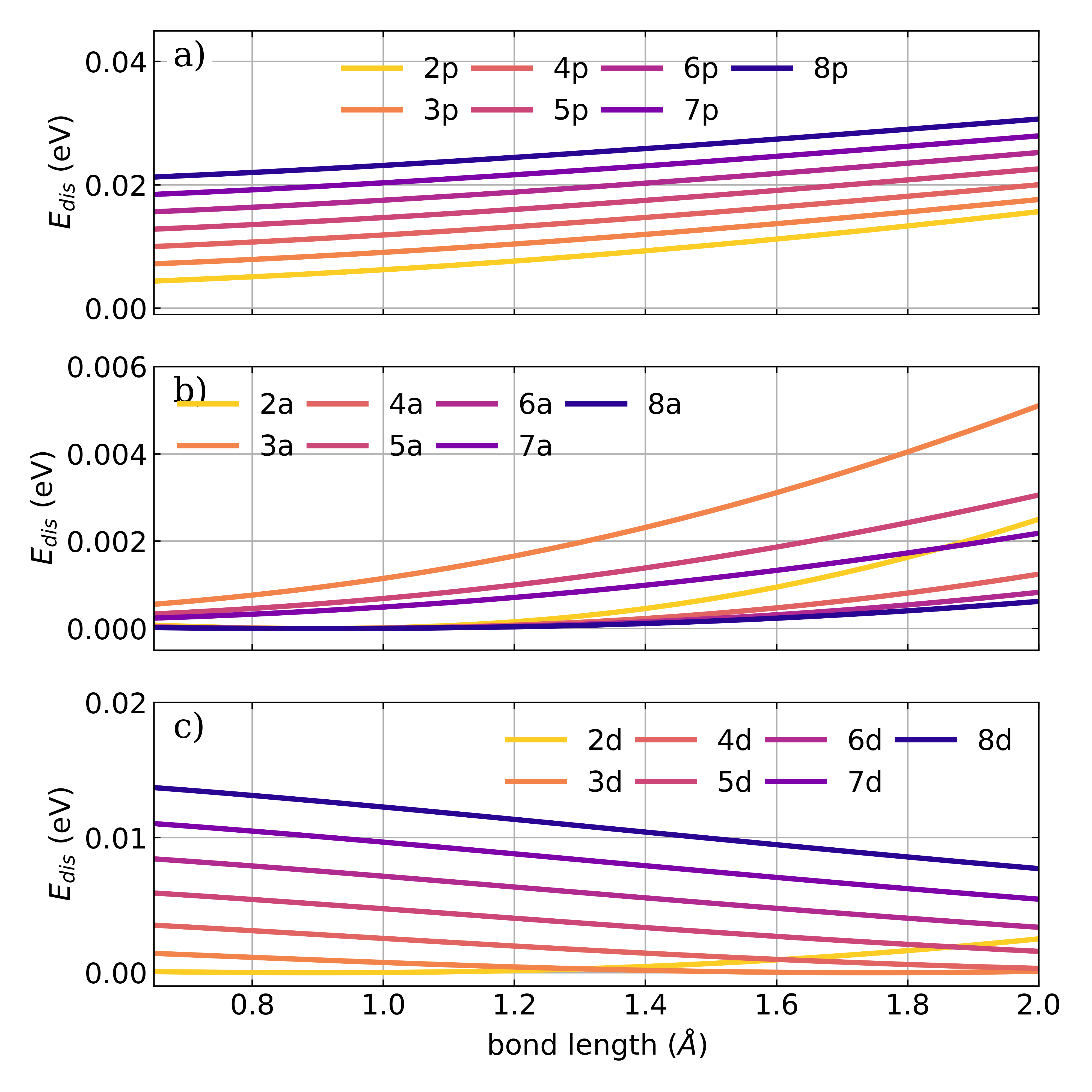}
    \caption{The cavity potential $E_{dis}$ along the \ce{HF} bond length for different ensemble sizes is shown for a) the \textit{all-parallel} configuration, b) the \textit{antiparallel} configuration and c) the \textit{defective} configuration. A cavity frequency $\omega_c$ of \SI{4467}{\per\centi\meter} is used. The strength of the cavity field $\epsilon_{c}$ is \SI{1.5}{\volt\per\nano\metre} and the number of molecules in the ensemble is color-coded.The used coupling strength $\bm{\lambda}_{c}$ is rescaled according to Eq.~16.}
\label{fig:hf_scan_vp}
\end{figure}

\begin{figure}[htb!]
     \centering
         \includegraphics[width=0.8\textwidth]{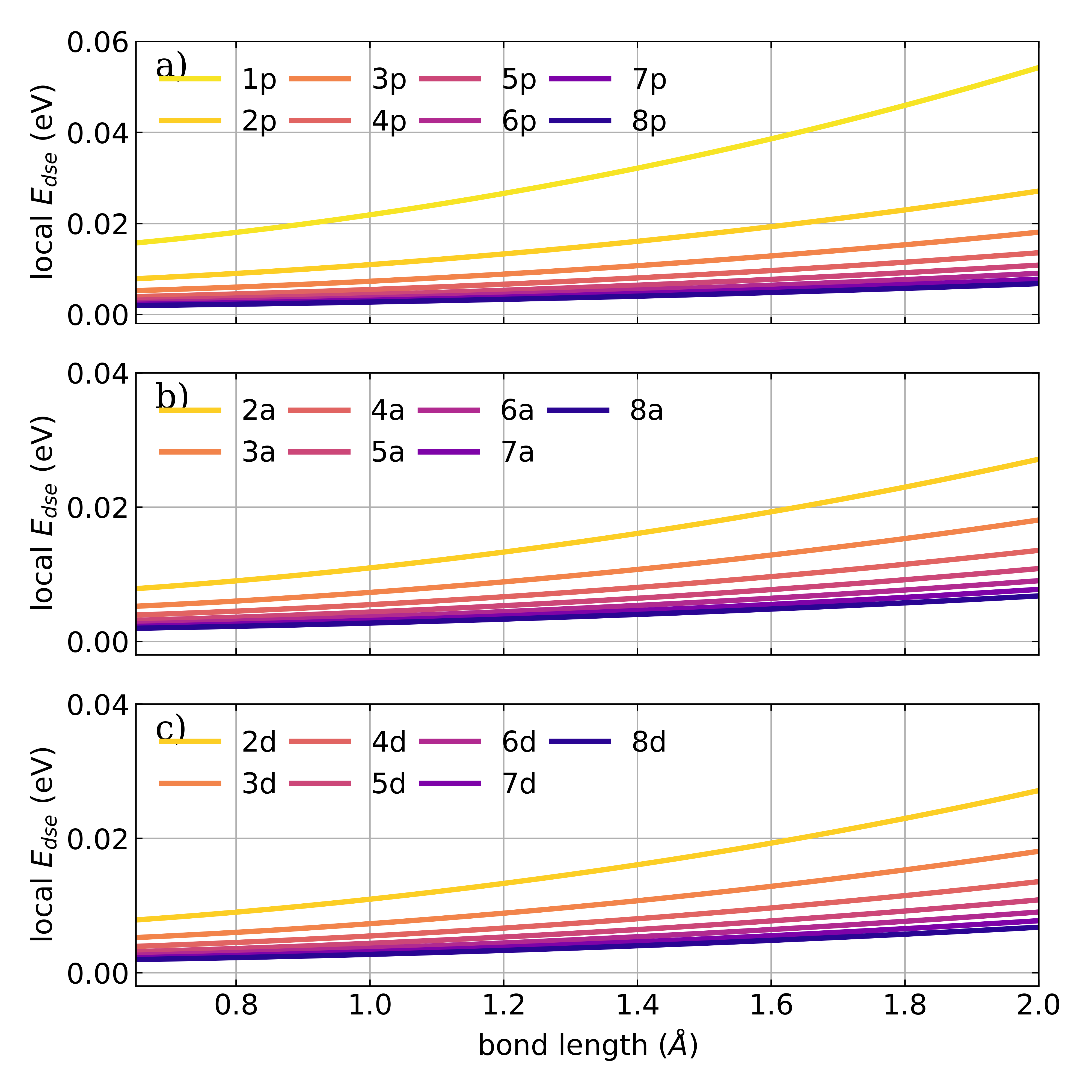}
    \caption{The local part of $E_{dse}$  along the \ce{HF} bond length for different ensemble sizes is shown for a) the \textit{all-parallel} configuration, b) the \textit{antiparallel} configuration and c) the \textit{defective} configuration. A cavity frequency $\omega_c$ of \SI{4467}{\per\centi\meter} is used. The strength of the cavity field $\epsilon_{c}$ is \SI{1.5}{\volt\per\nano\metre} and the number of molecules in the ensemble is color-coded.The used coupling strength $\bm{\lambda}_{c}$ is rescaled according to Eq.~16.}
\label{fig:hf_scan_dseL}
\end{figure}

\begin{figure}[htb!]
     \centering
         \includegraphics[width=0.8\textwidth]{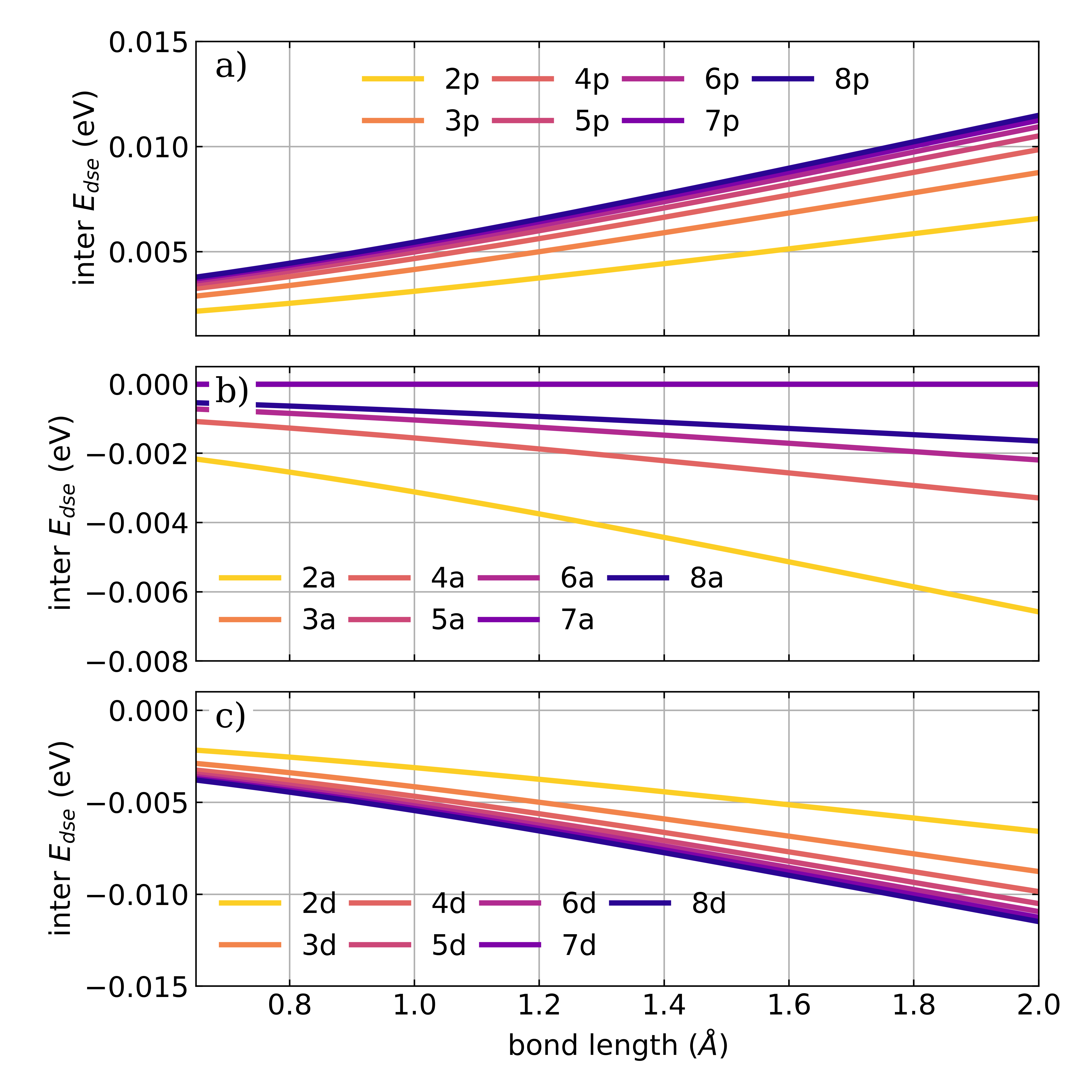}
    \caption{The local interaction part of $E_{dse}$ along the \ce{HF} bond length for different ensemble sizes is shown for a) the \textit{all-parallel} configuration, b) the \textit{antiparallel} configuration and c) the \textit{defective} configuration. A cavity frequency $\omega_c$ of \SI{4467}{\per\centi\meter} is used. The strength of the cavity field $\epsilon_{c}$ is \SI{1.5}{\volt\per\nano\metre} and the number of molecules in the ensemble is color-coded.The used coupling strength $\bm{\lambda}_{c}$ is rescaled according to Eq.~16.}
\label{fig:hf_scan_dseI}
\end{figure}

Supplementary results for ensemble energy changes along the scan of a single \ce{HF} bond are shown in Figs.~\ref{fig:hf_scan_part_para_ens}, \ref{fig:hf_scan_part_anti_ens}, and \ref{fig:hf_scan_part_dis_ens}. Individual energy contributions for the dissociating \ce{HF} molecule in different ensembles are shown in Figs.~\ref{fig:hf_scan_linear}, \ref{fig:hf_scan_vp}, \ref{fig:hf_scan_dseL}, and \ref{fig:hf_scan_dseI}.

\clearpage

\bibliography{lit.bib}